\documentclass{article}
\usepackage[margin=2cm]{geometry}
\usepackage[affil-it]{authblk}
\usepackage[utf8]{inputenc}
\usepackage{nccmath}
\usepackage{color}
\usepackage{tikz}
\usepackage{keyval}
\usepackage{ifthen}
\usepackage{epsf}
\usepackage{psfrag}
\usepackage{wrapfig}
\usepackage{color,rotating,psfrag,amsfonts}
\usepackage[dvips]{epsfig}
\usepackage{adjustbox}

\usepackage[backend=bibtex,style=numeric, maxbibnames=7, minbibnames=7, natbib=true, giveninits=true]{biblatex}
\usepackage{graphicx} 
\usepackage{xcolor}
\usepackage{float}
\usepackage{subfig}%
\usepackage{amsfonts,amsmath,amssymb,amsthm}
\usepackage{booktabs} 
\usepackage{overpic}

\usepackage{stackengine}
\stackMath
\newcommand\tsup[2][2]{
 \def\useanchorwidth{T}
  \ifnum#1>1
    \stackon[-1.3ex]{\tsup[\numexpr#1-1\relax]{#2}}{\mathchar"307E}
  \else
    \stackon[-1ex]{#2}{\mathchar"307E}
  \fi
}

\usepackage{pgfplots}
\usepackage{scalefnt}
\usetikzlibrary{arrows.meta}
\tikzset{>={Latex[round, length=1mm,width=0.5mm]}}
\usetikzlibrary{patterns,perspective}

\usepackage[utf8]{inputenc}
\usepackage{amsmath}
\usepackage{calc}

\usepackage{accents}

\usepackage{mleftright}

\newcommand{\co}{CO$_2$}
\newcommand{\ysamp}{Y^{(n)}}
\newcommand{\xsamp}{X^{(n)}}

\newcommand{\ctrl}{r}
\newcommand{\nctrl}{N_{\ctrl}}
\newcommand{\perm}{X}
\newcommand{\figref}[1]{Figure \ref{#1}}

\newcommand{\secref}[1]{Section \ref{#1}}

\newcommand{\tableref}[1]{Table \ref{#1}}

\newcommand{\Na}{N_{\text{a}}}



\title{Risk sharing in cooperative game models for CO$_2$ storage with uncertain geology and pressure competition}

\date{}

\author[1]{Per Pettersson\thanks{Corresponding author: per.pettersson@norceresearch.no}}
\author[1]{Svenn Tveit}
\author[1,2]{Sarah Gasda}

\affil[1]{NORCE Norwegian Research Centre, N-5838 Bergen, Norway.}
\affil[2]{University of Bergen, N-5020 Bergen, Norway}

\begin{document}

\maketitle

\begin{abstract}
With an increasing number of prospective geological CO$_2$ storage projects and potential pressure communication between different projects,  risk sharing under uncertain geological conditions are relevant to many project operators. In this work, the project operators are modeled as agents in a stochastic cooperative game. The agents can have different risk attitudes, here defined as being willing to accept more or less uncertainty in the predicted storage of CO$_2$. This uncertainty stems from lack of knowledge of geological parameters as well as unknown future actions of competing agents, and the corresponding probability distributions need to be estimated by numerical simulation.

The agents can choose  to share commercial risk if collaboration is preferable to a baseline scenario of individual injection. If their operations affect each other by means of, e.g., pressure communication, there may be no unique natural definition of a baseline scenario. As a remedy, we suggest belief distributions that combine uncertainty in physical data with maximum entropy prior distributions over the sets of viable injection actions.

For a realistic storage site, exemplified by the Utsira Formation in the North Sea, we present numerical results for both cases of pressure competition, and no hydraulic connections between different project operations. It is shown that risk averse agents benefit from collaboration when there is no pressure communication or other interference between agents. It is also demonstrated that pressure communication leads to large variability in the feasible injection rates, but the resulting belief distributions are nevertheless informative and useful for decision making about collaboration.

\end{abstract}

\section{Introduction}
The Intergovernmental Panel on Climate Change estimates that permanent storage of 3-10 Gt  CO$_2$ per year needs to be realized by the year 2050 to limit global warming to 1.5 degrees~\cite{IPCC_23}. The increase in industrial scale CO$_2$ storage operations required to reach these goals would be historically unique~\cite{Zahasky_Krevor_20} and is contingent on the exploitation of large regional aquifers by multiple storage projects~\cite{deJongeAnderson_etal_26}. On a global basis, regional aquifers are ubiquitous and have abundant pore volume, but the scale-up from single to multiple sites in simultaneous operation comes with new challenges~\cite{Ougier-Simonin_etal_26}. Pressure interference between projects will result in an aggregated pressure build-up across a wide region. In most cases, cumulative injected volumes need to be limited to maintain the geomechanical integrity of the caprock and faults. As a result, the risk of exceeding pressure limits or having to reduce injection rates to maintain storage integrity becomes a challenge that involves several independent actors, and their related commercial decisions, interacting over large spatial scales.

Risk assessment and management are common features of industrial-scale CO$_2$ storage. The subject of geomechanical and storage integrity risk has been addressed by operating projects such as Quest~\cite{Shell_10} and Decatur~\cite{Senel_Chugunov_13}, albeit the topic remains an active area of research \cite{Pettersson_etal_22, Goertz-Allmann_etal_24, Gasanzade_etal_26}. Commercial risk related to investment decisions at this scale are also handled routinely by industry. Here, the basic premise is to estimate the probability of a return on investment given the uncertainties inherent to the project, i.e. the value of the \co\ stored after appropriate cost deductions. The risk of falling short reduces the return and possibly incurs additional cost, e.g. drilling a new well or penalty for breach of contract. 
Geological uncertainties as well as external policies and market factors have a large impact on the expected outcome and its variability. These uncertainties are considered with respect to individual project's risk attitude in the final investment decision. To date, established practice for such commercial risk analysis has been refined with the context of single standalone projects. However, scale-up to multi-storage regions and the added impact of pressure competition expands the scope and complexity of each project's risk evaluation, which in turn challenges the conventional tools and analysis. 

Geological uncertainty can never be completely eliminated from commercial decisions. The oil and gas industry commonly handles this problem through risk sharing, i.e. spreading risk across a large portfolio of producing fields, which is analogous to a investing in a portfolio of stocks. Oil and gas is a mature industry, but the concept can be adapted for CO$_2$ storage to a certain degree, accounting for the geological uncertainty of storage projects that is often much larger and the commercial risk evaluation that is based on an emerging market with lower margins.  As the number of geological CO$_2$ storage projects increases worldwide, there will be a growing need for project operators to collaborate on risk sharing through insurance pools~\cite{Hiebert_24}. However, the issue of pressure communication adds an interdependency and complexity that is absent in the oil and gas analogue. The regional-scale context therefore introduces a qualitative shift in how risk must be understood and managed. Therefore, it is necessary to adapt conventional tools to address the unique challenge of CO$_2$ storage companies sharing space in large regional aquifers.

The basic premise of risk sharing is for all involved agents to agree on a model for distributing payoffs from a pooled or shared value, which is apriori unknown. The term payoff is generic in game theory and does not necessarily imply a positive reward. Risk sharing can be modeled using cooperative games with uncertain payoffs~\cite{Suijs_etal_99}, a common tool in the insurance industry.  This class of risk sharing models was first introduced in~\cite{Charnes_Granot_73} and has since been extended and generalized to cover different types of agents, risk attitude, and prior knowledge, see e.g.,~\cite{Granot_77, Doan_Nguyen_20}. The choice of risk allocation is negotiable, and each agent will need to evaluate any proposed scheme according to their risk attitude. For example, risk-averse agents will choose a payoff with a higher degree of certainty. Ultimately, the statistical data used in negotiations is very much dependent on the application, but to date, these concepts have not been evaluated within a \co\ storage context. 
Risk sharing models to evaluate the benefits of one collaboration model over another, are dependent on quantitative data, usually statistics of past situations. The challenge with CO$_2$ storage is the lack of relevant historical data. Instead, numerical models informed by geological models and data are the most reliable prediction tools available. While numerical simulation under geological uncertainty is commonplace, the context of pressure competition adds another source of uncertainty, namely the decisions of other actors in the same shared space.

Game theory has previously been proposed as a tool for decision making and cost-benefit distribution among independently acting agents sharing subsurface resources. Non-cooperative games modeled as decision trees were introduced to model the preferred development of gas and hydrocarbon fields in~\cite{Willigers_etal_09}. The possibility of a joint hub were included in the model, thus including an element of cooperation among the agents. Cooperative games with payoffs from multi-objective optimization of CO$_2$ safely injected by different coalitions of agents on a hydraulically connected site were introduced and analyzed in~\cite{Pettersson_etal_25}. The situation was modeled as a coalition formation generation problem, where agents were assumed to collaborate within their coalition, and compete for pressure space between coalitions. That analysis demonstrated how pressure interference makes achievable injection volumes dependent on coalition structure, even when operating limits are fixed. Reinforcement learning and surrogate models were proposed in~\cite{Chen_Hosseini_25} to identify injection strategies, while in~\cite{HernandezMejia_Pyrcz_25} the total pressure contribution was distributed to a number of injection wells by means of Shapley values.

Achieving maximum use of storage resources requires joint coordination between the storage project operators, either by voluntary means or incentivized by proper legislation. In a game theoretic setting, they would tentatively form the \textit{grand coalition}, containing all agents, and set the injection rates per well in order to optimize the total storage capacity of the regional aquifer. It is however an open question whether the value obtained by this grand coalition can be distributed by means of payoffs to all agents so that they are all satisfied, and have no incentive to deviate from collaboration. Whether such a stable solution exists at all needs to be investigated on a case-by-case basis.

In this paper, we propose a new way to evaluate the benefits of risk sharing between independent storage projects operating in a shared regional aquifer.  As a result, risk sharing decisions can no longer be evaluated independently: the feasibility and stability of a risk-sharing agreement depend on how pressure interference impacts or redistributes the variability in injected volumes across co-existing projects.  To solve this challenge, the application of cooperative games has to link geological uncertainty, pressure interaction, and uncertainty in collaboration decisions. 
Throughout the paper, the definition of risk is the uncertainty in the value of the injected CO$_2$ earned by each project. 
This is a generic definition of risk, and  we emphasize that in certain contexts it may be of interest with a more specific definition, such as risk being defined as the probability of an unusual and unwanted extreme event~\cite{Pettersson_etal_22}.

The remainder of the manuscript is structured as follows. In Section~\ref{sec:math-mod}, we present the mathematical model for decision making when the options are described by random variables, i.e., the outcomes are unknown, but described by well-defined probability distributions.
We then describe how risk sharing can be implemented by partitioning the total injection masses into stochastic payoffs for each agent, based on the agents' approach to risk and on how much they can expect to inject given the stochastic geological properties. In Section~\ref{sec:num-gen-data}, we briefly describe the physical model for the Utsira Formation in the North Sea, and its implementation with numerical optimization, to be used as an illustrative test case to demonstrate risk sharing in a realistic setting. Section~\ref{sec:num-no-comp} presents   numerical results for risk sharing when there is no pressure space competition, followed by a case when there is pressure competition in Section~\ref{sec:num-press-comp}. The variability due to uncertain geology and uncertain behavior of the individual agents is highlighted, and we propose two belief distribution models for the case of full or partial competition, such that all agents can make an informed decision about whether they should choose to collaborate. Numerical results for this case are then presented. Finally, we provide a discussion in Section~\ref{sec:discussion} and end with conclusions in Section~\ref{sec:conclusions}.

\section{Stochastic cooperative game models for risk sharing}
\label{sec:math-mod}

We consider a set of $\Na$ agents denoted $a_1, a_2,\dots, a_{\Na}$, where each agent is an independently-operated storage project owned by a company or business partnership. Depending on pressure communication, an agent may be in competition with the other agents.
Each agent injects {\co} over a period of time, and the total stored {\co} aggregated over all agents is taken as the total value, denoted $Y$. The agents can choose to form coalitions for their mutual benefit, where a coalition is a subset of agents that are assumed to coordinate their actions for the benefit of the group as a whole. A given partition of all of the agents into coalitions is referred to as a coalition structure. Of particular interest is the grand coalition, where all agents form a single joint coalition for mutual collaboration. If the agents indeed choose to form the grand coalition, the total value $Y$ will be distributed among the agents. Computational game theory offers a tractable way of evaluating the cost-benefits of entering into a risk-sharing coalition by means of coalition formation generation to identify the optimal coalition structure among a potentially overwhelmingly large set of different options~\cite{Sandholm_etal_99}. 

 In a deterministic setting, i.e. zero risk, the grand coalition is stable if all agents receive a share of the total value so that they do not prefer to leave the grand coalition. In this case, it is typically straightforward to determine whether an agent prefers a given outcome over another by simply comparing the respective payoffs. However, in the case of uncertainty, each agent's payoff will always be a random variable whether they choose to collaborate or not. Uncertainty makes the comparison of different stochastic payoffs non-trivial.
 
 Section~\ref{sec:pref-rel} describes a practical method for preference evaluation, i.e. ranking uncertain payoffs that takes into account one's own risk attitude. Section~\ref{sec:leakage} describes how preference relations can simultaneously account for both positive and negative  outcomes, e.g. storage revenue and penalty for leakage. 
Section~\ref{sec:payoff-alloc-models} then describes  a model for distributing stochastic payoffs. Different reasonable choices are proposed and discussed. The final section~\ref{sec:coll_analysis} describes the overall set of conditions that need to be met for the collaboration to be stable and explore different aspects of how agents can negotiate more favorable terms.

 \subsection{Preference among uncertain alternatives}
\label{sec:pref-rel}

In the situation of risk sharing, agents are faced with a decision to enter into a collaboration, a choice that involves comparing stochastic payoffs. 
There is no unique natural and direct way to tell whether one random variable is more `attractive' than another. Not only are random variables difficult to compare, the perception of attractiveness is intrinsically linked to each particular agent's risk attitude. The purpose of this subsection is therefore to elaborate on this fact, and suggest the \textit{preference relation} as a method to make such choices. 

A preference relation provides a direct comparison between random
variables and is obtained by a function that maps the random variables (tentative payoffs) to real numbers
that can be directly compared, ideally in such a way that it reflects the underlying risk attitude. Risk attitude is often divided into three categories: risk averse, risk neutral, and risk seeking behavior. Faced with the choice between a random payoff and the expectation of that payoff, a risk averse agent will choose the expectation over the random payoff, while a risk-seeking agent will choose the random payoff over the expectation, and the risk neutral agent is indifferent between the two choices~\cite{Suijs_etal_99}.

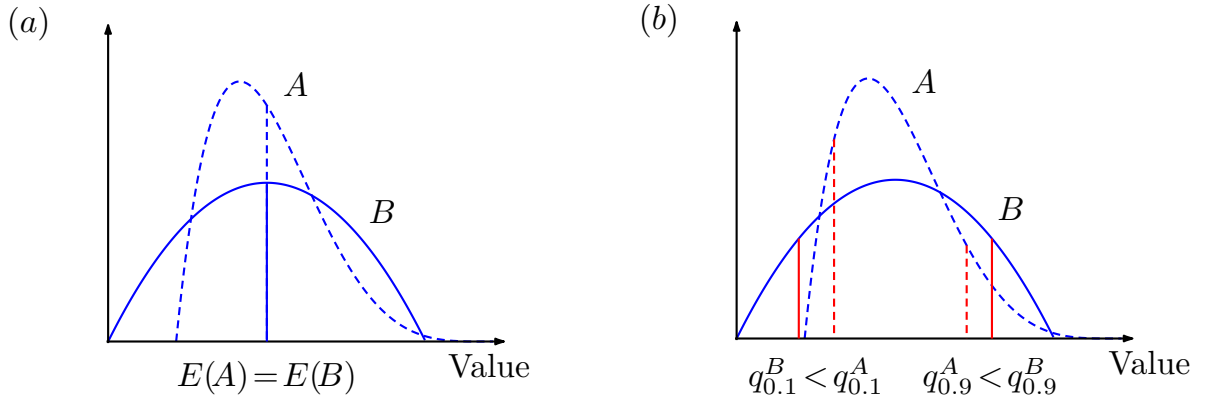
\begin{figure}[ht]
\tikzset{>={Latex[round, length=2mm,width=1.2mm]}}
\centering  
\subfloat 
{ 
\begin{tikzpicture}[thick,scale=4.2, transform shape] 

\draw (-0.05,1.0) node[left] {\scalebox{0.3}{$(a)$}};

     \def\boneshift{0.2143}
     \def\boneleft{0.0926}
     \def\btwoleft{0.1958}

     \def\boneright{0.5103}
     \def\btworight{0.8042}
     
    \def\betaone{\x,{(\x-\boneshift)*(1-(\x-\boneshift))^4*10}}
    \def\betatwo{\x,{\x*(1-\x)^1*2}}
     \def\fonemean{((0.5-\boneshift)*(1-0.5+\boneshift)^4)*10}

    \draw[color=blue, densely dashed, domain=\boneshift:1+\boneshift,samples=201] plot (\betaone) node[right] {};      
    \draw[color=blue,domain=0:1, samples = 201] plot (\betatwo) node[right] {};

    \def\ftwomean{(0.5*(1-0.5)^1)*2}

   \draw[->] (-0.005,0) -- (1.25,0) node[right] {};
    \draw[->] (0,-0.005) -- (0,1.0) node[above] {};
    
\draw (1.2,0) node[scale=0.3, below] {Value};

\draw (0.42,0.81) node[right] {\scalebox{0.3}{$A$}};
\draw (0.69,0.41) node[right] {\scalebox{0.3}{$B$}};

\draw[color=blue, densely dashed] ({0.5},{\fonemean}) -- ({0.5},0) node[below] {};

\draw[color=blue] ({0.5},{\ftwomean}) -- ({0.5},0) node[below] {};

\draw (0.5,0.07) node[below] {\scalebox{0.3}{$E  ( \!A  ) \! = \! E(\! B)$}};

\end{tikzpicture}
}
\subfloat 
{ 
\begin{tikzpicture}[thick,scale=4.2, every node/.style={transform shape}]

\draw (-0.05,1.0) node[left] {\scalebox{0.3}{$(b)$}};

     \def\boneshift{0.2143}
     \def\boneleft{0.0926}
     \def\btwoleft{0.1958}

     \def\boneright{0.5103}
     \def\btworight{0.8042}
     
    \def\betaone{\x,{(\x-\boneshift)*(1-(\x-\boneshift))^4*10}}
    
    \def\betatwo{\x,{\x*(1-\x)^1*2}}

     \def\foneleft{((\boneleft)*(1-(\boneleft))^4)*10}
     \def\foneright{((\boneright)*(1-(\boneright))^4)*10}
     
    \draw[color=blue, densely dashed, domain=\boneshift:1+\boneshift,samples=201] plot (\betaone) node[right] {};      
    \draw[color=blue,domain=0:1, samples = 201] plot (\betatwo) node[right] {};  

    \def\ftwoyleft{(\btwoleft*(1-\btwoleft)^1)*2}
    \def\ftwoyright{(\btworight*(1-\btworight)^1)*2}

   \draw[->] (-0.005,0) -- (1.25,0) node[right] {};
    \draw[->] (0,-0.005) -- (0,1.0) node[above] {};
    
\draw (1.3,0) node[scale=0.3, below] {Value};

\draw (0.42,0.81) node[right] {\scalebox{0.3}{$A$}};
\draw (0.69,0.41) node[right] {\scalebox{0.3}{$B$}};

\draw[color=red, densely dashed] ({\boneleft+\boneshift},{\foneleft}) -- ({\boneleft+\boneshift},0) node[below] {};
\draw[color=red, densely dashed] ({\boneright+\boneshift},{\foneright}) -- ({\boneright+\boneshift},0) node[below] {};
\draw[color=red] ({\btwoleft},{\ftwoyleft}) -- ({\btwoleft},0) node[below] {};
\draw[color=red] ({\btworight},{\ftwoyright}) -- ({\btworight},0) node[below] {};
\draw (0.25,0.07) node[below] {\scalebox{0.3}{$q_{0.1}^{B} \! < \! q_{0.1}^{A}$}};
\draw (0.8,0.07) node[below] {\scalebox{0.3}{$q_{0.9}^{A} \! < \! q_{0.9}^{B}$}};

\end{tikzpicture}
}
\caption{\label{fig:pref-rel-example} An agent can choose between random options $A$ and $B$ with the same mean, but otherwise different distributions.}
\end{figure}

Denote the preference by $\succeq$ so that for any two options $A$ and $B$ (described by random variables) that an agent can choose from, the relation $A \succeq B$ means that $A$ is preferred over $B$. For
instance, in the somewhat unrealistic case that risk is ignored and only expected payoffs matter, the mapping from any
random variable to its expectation (mean value), is a perfectly valid preference relation. The preference relation concept is illustrated in Figure~\ref{fig:pref-rel-example}, where an agent is posed with a choice between two random variables $A$ and $B$, which represent storage profits for the purposes of this example. The probability density functions (PDFs) of $A$ and $B$, happen to share the same mean value but follow otherwise different distributions. A few possible preferences can be immediately surmised. First, the expectation preference is a straightforward comparison of the mean values, but ignores the tail probabilities. We observe that option $A$ has zero risk for outcomes of small value, which may be attractive compared with non-zero probability for option $B$. On the other hand, $B$ offers higher probability of obtaining large values, and may hence be an attractive choice in that regard. Ultimately, there are multiple ways to compare $A$ and $B$ that have different properties. 
The preference is said to be complete if the agent can always say that among two random options, one is better than the other, or they are equally
good. The mean value preference is complete in this sense since we can always tell which one is greater of two mean
values, or whether they are equal. Some preference measures fail to provide some of these answers in all possible
situations, and are therefore called incomplete.

A suitable preference relation should, first and foremost, reflect what an agent actually prefers to be meaningful in practice. An agent that is risk averse would choose a low-risk option over a high-risk option, everything else being equal. In the current work, a suitable preference is defined by a given quantile of the random variables to be compared. For the example with options $A$ and $B$, the 0.1 and 0.9 quantiles are shown in Figure~\ref{fig:pref-rel-example}. An agent that wants to avoid negative outcomes (i.e. small values) would consider the 0.1 quantiles of $A$ and $B$, and based on this preference relation choose option $A$. Meanwhile, an agent that cares more about the probability of winning big would consider the 0.9 quantile as its preference relation, and choose $B$ over $A$. We will use the notation 
\begin{equation}
\label{eq:quantile-pref}
A \succeq_{\alpha} B \mbox{ if and only if } q_{\alpha}^{A} \geq q_{\alpha}^{B}, 
\end{equation}
with $q_{\alpha}^{A}$ being the $\alpha$-quantile ($0\leq \alpha \leq 1$) where $\alpha$ is the probability that the random variable $A$ is less than or equal to  $q_{\alpha}^{A}$. Since different agents can have different preferences, we sometimes use the notation $\succeq_{\alpha_i}$ to denote the quantile preference of agent $i$. An advantage of $\succeq_{\alpha}$ is that it is complete (i.e., always tells whether one option is better or equal to another), and can model different behavior with respect to risk, in the sense that a small $\alpha$ corresponds to risk averse behavior, and a large $\alpha$ corresponds to more risk seeking behavior. 

Other preferences include the already mentioned expectation preference, and stochastic dominance preference~\cite{Hadar_Russell_69}. The expectation preference does not distinguish between risk averse, neutral, and risk seeking behavior, and is therefore too crude to be of interest in the current paper. The stochastic dominance preference compares the full cumulative density functions, that is all possible quantiles.
This preference is incomplete, something that is not only an issue for some rare cases, but would be the case for several of the examples given later in this paper.
The shortcomings of the dominance preference is illustrated in Figure~\ref{fig:pref-rel-example}(b); since $q_{\alpha}^{B} < q_{\alpha}^{A}$ for $\alpha=0.1$, but $q_{\alpha}^{A} < q_{\alpha}^{B}$ for $\alpha=0.9$ it follows that neither $A$ nor $B$ is to be preferred according to the criterion, but $A$ and $B$ are also not equally preferable. Thus, this preference relation does not provide an answer (preferred, not preferred, or indifferent) for this example. However, using either quantile (or some other single quantile) gives a unique answer, demonstrating the applicability of the quantile preference.

\subsection{Including penalties in payoffs and risk preferences}
\label{sec:leakage}
\begin{figure}[ht]
\tikzset{>={Latex[round, length=2mm,width=1.2mm]}}

\centering  

\subfloat[Risk averse behavior.]  
{ 
{\scalefont{0.8}
\begin{tikzpicture}[thick,scale=1.3, every node/.style={transform shape}]
 
     \def\betaone{\x,{(\x/2)*(1-\x/2)^4*30/3*2}}
     \def\betatwo{\x,{\x/2*(1-\x/2)^1*2*2}}
   
     \def\yone{1.0206}
     \def\ytwo{1.6084}
     
    \def\foney{(\yone/2*(1-\yone/2)^4)*20}
    \draw[color=blue, densely dashed, domain=0:2,samples=201] plot (\betaone) node[right] {};  
    \draw[color=red, densely dashed] ({\yone},{\foney}) -- ({\yone},0) node[below] {};

    \draw[color=blue,domain=0:2, samples = 201] plot (\betatwo) node[right] {};  
    \def\ftwoy{(\ytwo*(1-\ytwo)^1.5)*8.7489/3}
    \def\ftwoy{(\ytwo*(1-\ytwo)^0.5)*3.75/3}
    \def\ftwoy{(\ytwo/2*(1-\ytwo/2)^1)*2*2}

    \draw[color=red] ({\ytwo},{\ftwoy}) -- ({\ytwo},0) node[below] {};
    
   \draw[->] (-0.005,0) -- (2.4,0) node[right] {};
    \draw[->] (0,-0.005) -- (0,2.0) node[above] {};
    
    \draw (2.3,0) node[below] {$L_{i}$};
    
\draw (1.3,0) node[below] {$q_{\alpha'}^{A} \! < \! q_{\alpha'}^{B}$};

\draw (0.52,1.62) node[right] {$A$};
\draw (1.45,0.85) node[right] {$B$};

    \draw (2.5,1.2) node {$\alpha = 1-\alpha'$};
     \draw (2.5,1.0) node {$\Longrightarrow$};
    
    \def\yone{3.9794}
     \def\ytwo{3.3916}
    \def\betaone{\x,{(5-\x)/2*(1-(5-\x)/2)^4*30/3*2}}
    \def\foney{((5-\yone)/2*(1-(5-\yone)/2)^4)*20}
    
    \def\betatwo{\x,{(5-\x)/2*(1-(5-\x)/2)^1*2*2}}
    \def\ftwoy{((5-\ytwo)/2*(1-(5-\ytwo)/2)^1)*4}

    \draw[color=blue, densely dashed, domain=3:5,samples=201] plot (\betaone) node[right] {};  
    \draw[color=red, densely dashed] ({\yone},{\foney}) -- ({\yone},0) node[below] {};
    
    \draw[color=blue, domain=3:5,samples=201] plot (\betatwo) node[right] {};  
    \draw[color=red] ({\ytwo},{\ftwoy}) -- ({\ytwo},0) node[below] {};
    
   \draw[->] (2.8,0) -- (5.3,0) node[right] {};
    \draw[->] (5.0,-0.005) -- (5.0,2.0) node[above] {};
    
    \draw (5.3,0) node[below] {$-L_i$};
    
    \draw (3.68,0) node[below] {$q_{\alpha}^{B} <  q_{\alpha}^{A}$};

    \draw (4.0,1.62) node[right] {$A$};
    \draw (3.1,0.85) node[right] {$B$};

\end{tikzpicture}
}
}
\subfloat[Risk seeking behavior.]  
{ 
{\scalefont{0.8}
\begin{tikzpicture}[thick,scale=1.3, every node/.style={transform shape}]

     \def\betaone{\x,{(\x/2)*(1-\x/2)^4*30/3*2}}
     \def\betatwo{\x,{\x/2*(1-\x/2)^1*2*2}}
   
     \def\yone{0.2244}
     
    \def\foney{(\yone/2*(1-\yone/2)^4)*20}
    \draw[color=blue, densely dashed, domain=0:2,samples=201] plot (\betaone) node[right] {};  
    \draw[color=red, densely dashed] ({\yone},{\foney}) -- ({\yone},0) node[below] {};
 
    \draw[color=blue,domain=0:2, samples = 201] plot (\betatwo) node[right] {};  
    \def\ftwoy{(\ytwo/2*(1-\ytwo/2)^1)*2*2}
    
    \def\ytwo{0.3916}
    
    \draw[color=red] ({\ytwo},{\ftwoy}) -- ({\ytwo},0) node[below] {};
    
   \draw[->] (-0.005,0) -- (2.4,0) node[right] {};
    \draw[->] (0,-0.005) -- (0,2.0) node[above] {};
    
    \draw (2.3,0) node[below] {$L_{i}$};
    
\draw (0.3,0) node[below] {$q_{\alpha'}^{A} \! < \! q_{\alpha'}^{B}$};

\draw (0.52,1.62) node[right] {$A$};
\draw (1.45,0.85) node[right] {$B$};

\draw (2.5,1.2) node {$\alpha = 1-\alpha'$};    
\draw (2.5,1.0) node {$\Longrightarrow$};
    
    \def\yone{4.7756}
     \def\ytwo{4.6084}
    
    \def\betaone{\x,{(5-\x)/2*(1-(5-\x)/2)^4*30/3*2}}
    \def\foney{((5-\yone)/2*(1-(5-\yone)/2)^4)*20}
    
    \def\betatwo{\x,{(5-\x)/2*(1-(5-\x)/2)^1*2*2}}
    \def\ftwoy{((5-\ytwo)/2*(1-(5-\ytwo)/2)^1)*4}

    \draw[color=blue, densely dashed, domain=3:5,samples=201] plot (\betaone) node[right] {};  
    \draw[color=red, densely dashed] ({\yone},{\foney}) -- ({\yone},0) node[below] {};
    
    \draw[color=blue, domain=3:5,samples=201] plot (\betatwo) node[right] {};  
    \draw[color=red] ({\ytwo},{\ftwoy}) -- ({\ytwo},0) node[below] {};
    
   \draw[->] (2.8,0) -- (5.5,0) node[right] {};
    \draw[->] (5.0,-0.005) -- (5.0,2.0) node[above] {};
    
    \draw (5.6,0) node[below] {$-L_i$};
    
    \draw (4.68,0) node[below] {$q_{\alpha}^{B} \! < \! q_{\alpha}^{A}$};

    \draw (4.1,1.62) node[right] {$A$};
    \draw (3.2,0.85) node[right] {$B$};

\end{tikzpicture}
}
}
\caption{Risk behavior in terms of quantile preferences can be applied to negative payoffs such as leakage, or failure to inject.}
\label{fig:risk-leakage}
\end{figure}
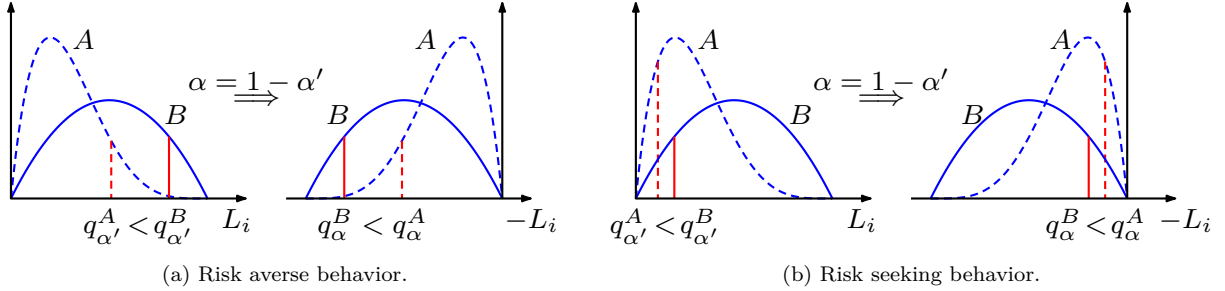

The generic definition of risk as the variability of the positive payoff can also include the effects from unwanted penalties, such as failure to inject according to contract, or leakage of CO$_2$. It is safe to assume that any agent will want the uncertain loss to be as small as possible. It may be less clear what it means to be risk averse or risk seeking in terms of loss, and also how the risk preference can handle loss and gains simultaneously. The purpose of this section is to make these two points more clear. As an illustration, consider two choices, as before labeled $A$ and $B$, that lead to different probability distributions for monetary loss or leakage $L_i$ for agent $a_i$. The choice $A$ has a PDF that should intuitively make it the preferred choice over $B$ independent of actual preference relation, as shown in Figure~\ref{fig:risk-leakage}. Under the quantile preference relation, it is natural to assume that a risk averse agent will want to avoid more extreme losses rather than just making sure that it is likely that only a small amount will not be safely stored. Hence, for risk averse agents, a preference based on a quantile $\alpha' > 0.5$ captures this situation. However, $q^A_{\alpha'} < q^B_{\alpha'}$, which does not agree with the conventional use of the quantile preference for situations where clearly we would expect $A \succeq B$. In practice, any positive reward will be weighted against the negative loss $-L_i$. This corresponds to the mirrored PDF with $\alpha = 1-\alpha'$ for negative loss, and is illustrated in Figure~\ref{fig:risk-leakage}~(a). Thus, $q_{\alpha}^{A} > q_{\alpha}^{B}$ with $\alpha<0.5$, as desired. 

A risk seeking agent will of course also want to avoid leakage. However, one may argue that a truly risk seeking agent may be content to make a choice where some limited leakage has a small probability of occurring, rather than minimizing the risk of a more extreme leakage scenario. Again, setting $\alpha=1-\alpha'$ and considering the PDF of $-L_i$, we get the result that an intuitively risk seeking attitude towards loss can be encoded with a quantile preference for large values of $\alpha$, just in the case with positive rewards. This is illustrated in Figure~\ref{fig:risk-leakage}~(b), where $q_{\alpha}^{A} > q_{\alpha}^{B}$ for $\alpha > 0.5$.

In reality, a negative loss will of course only be acceptable if it is compensated by a sufficient amount of positive payoff in terms of stored CO$_2$. If the cost of a non-stored unit of CO$_2$ is equal to $c_i$ units of safely stored CO$_2$ for $a_i$, we can express the total payoff $Y_i$ to agent $a_i$ in terms of stored ($S_i$) and lost CO$_2$ as follows:
\[
Y_i = S_i - c_i L_i,
\]
where $S_i$ and $L_i$ will be dependent random variables. Risk averse behavior with respect to $Y_i$ thus defined will simultaneously favor both avoidance of large leakage penalties and avoidance of only relatively small injected CO$_2$ volumes. In contrast, risk seeking behavior with respect to $Y_i$ favor the prospects of simultaneous small leakage penalties and large total injection volumes.
Although the numerical examples provided in this paper do not feature leakage, all theory presented is applicable to the above model for inclusion of leakage.

\subsection{Payoff distribution models}
\label{sec:payoff-alloc-models}
Now that we have introduced the basic elements for making choices between random variables, the next step is to consider different models for allocating risk, i.e. partitioning the total value into individual payoffs. If the agents decide to collaborate, the uncertain total payoff $Y$ should be distributed among them in accordance with their risk preferences. All agents will want a sufficiently large share of the total value CO$_2$ stored to enter a coalition, but an agent willing to take more risk should accept more uncertainty in its share compared to an agent that is risk averse. 
Essentially, the risk allocation model will return a new random variable that will be taken by each agent and compared against another one, as described above in Section~\ref{sec:pref-rel}. The risk allocation model is an \textit{a priori} model for pooling profits and redistributing them according to a pre-defined formula. This is similar to an insurance policy, where the payoff to each agent is still a random variable, but it will be different from an agent’s payoff if acting alone. We refer to this non-risk sharing arrangement as the baseline.

 Let $Y_i$ ($i=1,\dots, \Na$) denote the stochastic payoff (distributed value of injected CO$_2$) of agent $a_i$. Whether the agents inject in collaboration or not, the total amount $Y$ needs to be the sum of the individual contributions $Y_i$, i.e.,
\begin{equation}
\label{eq:Y_partition}
\sum_{i=1}^{\Na}Y_i = Y.
\end{equation}
We first consider the case of full collaboration, in which the agents act as a single entity and the uncertainty in the total value $Y$ is directly given by the stochastic physical model (only geological uncertainty). Hence, $Y$ can be regarded as some fixed random variable, whose distribution can be estimated by numerical simulation. In this case, there is some flexibility in the allocation of the total $Y$ into the individual payoffs $Y_i$, as long as the sum equals the total value. The exact random variable $Y_i$ is obtained through negotiation with the other agents. An individual agent will of course not accept an allocation that appears unfavorable with respect to what can be anticipated based on geological considerations, but the agent may for instance be willing to accept a higher risk in exchange for the prospect of higher expected injection volumes. The crucial point is that while there are physical constraints as well as economical considerations, there is typically no unique random $Y_i$ that agent $a_i$ finds acceptable, although some characteristics of $Y_i$ are more desirable than others.

In the following, we will discuss allocation of the total payoff $Y$ to the individual agents. We emphasize that in principle, the payoff allocated to an individual agent is not necessarily the amount of CO$_2$ that agent will inject. In fact, it may not be practical or even possible to inject exactly that amount of CO$_2$ in the wells operated by the agent in question. Instead, the individual allocation should be interpreted as a monetary value or credit corresponding to a well-defined and possibly non-trivial share of the total injection, established via contractual obligations in the pre-injection phase, and assuming that all agents work together to fulfill the joint goal. To make it clear that we will consider individual payoffs that are often related, but not necessarily equal, to $Y_i$, we will use notation with various superscripts to indicate that we consider re-allocation of resources. 

A deterministic reference value such as $E(Y)$, where $E(\cdot)$ denotes the expectation (mean value),
represents the agents' collective belief about the total value. The payoffs are then split into a deterministic component $d_i$, and a random component $r_i$ that is assumed to be a fraction  of $Y$, so that we get an allocation that is linear in $Y$, i.e. $Y_i=Y_i^{\text{lin}}$, where
\begin{equation}
\label{eq:payoff_split_Suijs}
Y_i^{\text{lin}} = d_i + r_i (Y-E(Y)), \quad i=1,\dots, \Na,
\end{equation}
with $r_i \geq 0$, and to ensure that Eq.~\eqref{eq:Y_partition} holds:
\[
\sum_{i=1}^{\Na}r_i =1, \quad  \mbox{and} \quad \sum_{i=1}^{\Na}d_i = E(Y).
\]
Note that the model both allows redistribution of risk ($r_i$) and transfer of deterministic payment $(d_i)$. The latter makes insurance agreements possible: one can trade risk against money, and this is a key feature of the current model~\cite{Borm_Suijs_02}.

We can now interpret the definitions of respectively risk averse, risk neutral, and risk seeking behavior described above in terms of the risk sharing parameters $d_i, r_i$. A risk averse agent will want $r_i$ to be small, and generally prefer a fixed deterministic payoff over an uncertain one. A risk neutral agent has no preference over the deterministic and the random component of its payoff, and a risk seeking agent prefers a larger $r_i$, i.e., a larger fraction of the random component of the payoff.
Note that in general the individual agent values -- if they are at all defined -- are not equal since well locations affect injection volumes and cost, so even if all agents have the same approach to risk, they should not have equal $d_i$, and not equal $r_i$.

While $(d_i, r_i)$ could be assigned any of a range of different values, some values may be more intuitive and merit further attention. Next, we will suggest three possible risk allocations. To this end, assume that there is a natural individual baseline
  allocation $Y_i=Y_i^{\text{ind}}$, that will be realized when there is no collaboration. Two of the three suggested payoff allocations will rely on the existence of $Y_i^{\text{ind}}$.
  
\textit{Proportional payoff:} For instance, $Y_i^{\text{ind}}$ can be the payoff when $a_i$ injects independently and unaffected by other agents in a scenario where no pressure interference effects can occur. In this case, Eq.~\eqref{eq:Y_partition} is satisfied with the terms being equal to the actual physical injections of each agent. With the choice of deterministic coefficients $d_i = E\mleft(Y_i^{\text{ind}}\mright)$ and $r_i = E\mleft(Y_i^{\text{ind}}\mright)/E(Y)$ in~\eqref{eq:payoff_split_Suijs},  we obtain a payoff proportional to each agent's expectation:
\begin{equation}
\label{eq:risk-sharing-phys-prop}
Y_i^{\text{prop}}
= E\mleft(Y_i^{\text{ind}}\mright) + \frac{E\mleft(Y_i^{\text{ind}}\mright)}{E(Y)}\left(Y-E(Y)\right)
=
\frac{E\mleft(Y_i^{\text{ind}}\mright)}{E\mleft(Y\mright)}Y.
\end{equation}
This is a natural choice since $E\mleft(Y_i^{\text{prop}}\mright)=E\mleft(Y_i^{\text{ind}}\mright)$, so each agent's expected redistributed payoff coincides with its expected physical payoff. Since $\sum_{i} Y_i^{\text{prop}} = Y$, this allocation scheme can be applied simultaneously to all agents. 

\textit{Full variability payoff: }Assuming unchanged expectations (i.e., $d_i=E\mleft(Y_i^{\text{ind}}\mright)$) for all agents, one agent can take a higher risk provided that the other agents prefer to take a lower risk. In the extreme cases, $r_i=0$ implies that $a_i$ will obtain its expected payoff with certainty, whereas $r_i=1$ means that $a_i$ accepts the full total risk, as given by the full range of the random distribution of $Y$, but still with its expectation $E\mleft(Y_i^{\text{ind}}\mright)$ unaffected. With $r_i=1$, we get the allocation
\[
Y_i^{\text{full}} 
= E\mleft(Y_i^{\text{ind}}\mright) + Y-E(Y).
\]

\textit{Cost-based payoff: }A third alternative is to assume that the agents want to get a payoff that is proportional to their investment costs, denoted $c_i$. Then, a possible allocation is:
\begin{equation}
\label{eq:risk-sharing-cost-prop}
Y_{i}^{\text{cost}} =
\frac{c_i}{\sum_{k=1}^{\Na} c_k}Y.
\end{equation}


\subsection{When is collaboration attractive?}
\label{sec:coll_analysis} 
If there exists a baseline or non-competitive scenario $Y_i^{\text{ind}}$ ($i=1,\dots, \Na$) with well-defined individual injection schedules, then a stable grand coalition risk sharing allocation $Y_i^{\text{lin}}$ in the form~\eqref{eq:payoff_split_Suijs}  implies that the following condition is satisfied:
\begin{equation}
\label{eq:mut_ben_crit}
Y_{i}^{\text{lin}} \succeq Y_i^{\text{ind}}, \quad i=1,\dots,\Na, \quad \mbox{(mutually beneficial)}.
\end{equation}
This means that all agents prefer the suggested payoff over their baseline payoff. We stress that the consistency criterion~Eq.~\eqref{eq:Y_partition}
is automatically satisfied via Eq.~\eqref{eq:payoff_split_Suijs}.

Ideally, one would like to find the optimal sets of $d_i$ and $r_i$ so that all agents obtain the best possible payoff. This is however not always possible in practice. It can still be helpful for agents in a situation of negotiation to map out what combinations of $d_i, r_i$ are equally attractive.
An agent $a_i$ that is risk averse ($\alpha < 0.5$) will want a small $r_i$ and a large $d_i$ (mean). There may exist distinct allocations $\mleft(d_i^{(1)},r_i^{(1)}\mright)$ and $\mleft(d_i^{(2)},r_i^{(2)}\mright)$ for $a_i$ such that  $d_i^{(1)} < d_i^{(2)}$ and $r_i^{(1)} < r_i^{(2)}$  so that $a_i$ does not prefer the one allocation over the other. 
When is this the case? In the following, we will assume the quantile preference~\eqref{eq:quantile-pref} for all agents. By the definition of the quantile and the cumulative distribution functions,
\begin{equation}
\alpha_i = P(d_i+r_i(Y-E(Y)) \leq q_{\alpha_i}) =
P(Y \leq (q_{\alpha_i}-d_i)/r_i+E(Y))
=
F_{Y}((q_{\alpha_i}-d_i)/r_i+E(Y)),
\end{equation}
Hence,
\begin{equation}
\frac{q_{\alpha_i}-d}{r}+E(Y) = F^{-1}_{Y}(\alpha_i), 
\Leftrightarrow 
q_{\alpha_i} = r_i \underbrace{(F^{-1}_{Y}(\alpha_i) - E(Y))}_{\mbox{constant } a^{\alpha}_{i}} + d_i = d_i + r_i a^{\alpha}_{i}. \notag
\end{equation}
Given an agent with a set of choices resulting in different combinations of two different quantities, the \textit{indifference curves} are graphs that indicate the values for which the agent is indifferent~\cite{Edgeworth_1881}. Here, the indifference curves are
the graphs of combinations of $(d_i, r_i)$ that yield the same quantile, i.e., $q_{\alpha_i} = d_i + a_i^{\alpha} r_i = \text{const}$. For total payoff distributions with expectation equal to the median, the indifference curves will always have positive slope for risk averse agents, and negative slope for risk seeking agents.

 Roughly speaking, in a setting with risk averse agents only, collaboration will then be attractive if the variance of the total payoff is small compared to the baseline payoffs of the agents, assuming fixed expectations. This will now be made more concrete.
In addition to the existence of equally preferable but different allocations, there may be allocations that are more preferable to all agents than their baseline individual payoffs $Y_i^{\text{ind}}$. If that is the case, all agents should be willing to collaborate and be satisfied with any of these allocations, although the prospect of an even better outcome could lead to disagreement about the exact distribution of payoff. As derived in Appendix~\ref{sec_appendix_redist}, more attractive allocations than $Y_i^{\text{ind}}$ exist if:
\begin{equation}
\label{eq:surplus_cond}
q_{\alpha}^{Y} > \sum_{i} q_{\alpha}^{Y_i^{\text{ind}}}.
\end{equation}
Hence, an easy-to-check condition for finding out when collaboration should take place is given by~\eqref{eq:surplus_cond}, which implies that~\eqref{eq:mut_ben_crit} holds.

There are some idealized but representative
situations where it is possible to make general conclusions about collaborations without the need to know the specific details about the distributions. We will give three examples, all assuming the proportional allocation~\eqref{eq:risk-sharing-phys-prop}.

As a first example, consider the idealized situation where the $Y_i^{\text{ind}}$s are independent and follow the same arbitrary distributions with non-zero variance. In this case, collaboration with the proportional allocation~\eqref{eq:risk-sharing-phys-prop} is always more attractive than no collaboration under the risk averse quantile preference measure. Likewise, collaboration is never attractive for a risk seeking agent. A proof is provided in Appendix~\ref{sec:appendix_IID}. Next, if the individual injections follow the same distribution and have perfect (positive) correlation, risk averse agents will still want to collaborate for decreased individual risk, see proof in Appendix~\ref{sec:appendix_ID_pos_corr}. Of course, any realistic situations will not display identical distributions, but as we will see in the numerical results section, this is a decent approximation. Also, although the strict proofs require identical distributions, the conclusion still holds in many practical situations of vaguely similar but not identical distributions.

As a third example, assume that all $Y_i^{\text{ind}}$ are independent as in the first example and have the same expectation, and that a subset of them denoted $J$ have some smaller variance determined by a factor $\epsilon < 1$, specifically $V\mleft(Y_j^{\text{ind}}\mright)=\epsilon V(Y_i^{\text{ind}})$ for $j \in J$ and $i \notin J$. Clearly, as in the first example, any risk averse agent $a_i$ with $i \notin J$ will still want to share the now reduced total risk. Less intuitive, any agent $a_j$ with $j \in J$ will want to collaborate only if $\epsilon > (\Na - |J|)/(\Na^2 - |J|)$, where $|J|$ denotes the number of agents with the smaller variance, so it is possible to obtain an exact value of how small the variance can be without changing the attractiveness of collaboration. In particular, with only one agent with smaller variance, collaboration is attractive only if $V\mleft(Y_j^{\text{ind}}\mright)>V\mleft(Y_i^{\text{ind}}\mright)/(\Na+1)$ for $i \notin J$. This means that we can give a quantitative answer to how large the variance of an individual agent needs to be to motivate entering a collaboration for risk sharing.  The proofs are provided in Appendix~\ref{sec:appendix_small_var}.


\section{Numerical generation of payoff probability density functions}
\label{sec:num-gen-data}
Next we describe the steps to obtain estimates of the probability density functions of the different payoff options. There are in general no direct data on storage capacity that are rich enough to be used for accurate PDF estimates. Instead, we assume a random field model for the permeability field of a representative storage site and propagate the uncertainty through a multi-phase flow model with fracture pressure constraints.   Uncertainty in other model properties, e.g., end-point relative permeabilities, fault locations and sedimentary layer boundaries, could also be included within the same computational framework.

\subsection{Monte Carlo simulations of optimal storage capacity}
\label{sec:MC-res-model}

Let the uncertainty in geological properties be encoded by a vector of random variables $X=(X_1,\dots, X_m) \in \mathbb{R}^{m}$, for some sufficiently large number $m$ of random variables. Complemented with a correlation model, $X$ can define e.g. a spatially varying permeability field via a Karhunen-Loeve expansion~\cite{Ghanem_Spanos_91}. The payoff (i.e., total amount of CO$_2$) $Y$ is then a random variable as a consequence of being a function of the physical model $M$, which is the maximum storage of a reservoir model $\tilde{M}$ subject to constraints $g$. The physically constrained model is a function of both the control variables $\ctrl$ and the random vector $X$:
\begin{equation}
    Y = M(X) = \max_{\ctrl} \tilde{M}(X; \ctrl),\quad\text{s.t.} \, g(X; \ctrl)\leq 0.
\label{eq:optim_prob}
\end{equation}

In this paper, the control variables $\ctrl$ are the temporally constant injection rates for each well. The constraint $g(\ctrl)$ is the maximum sustainable overpressure, $\Delta P_{\text{crit}}$, introduced to ensure that the maximum storage capacity is physically meaningful. We follow the approach in \cite{GASDA2017} and define $g=\Delta P_{\text{crit}}=P_{\text{frac}} - P_{\max}$ where $P_{\text{frac}}$ is the fracture pressure threshold of the rock, and $P_{\max}$ is the maximum pressure at any spatial location during the injection period. 
We define $Y$ as the maximum capacity [Mt] of the storage area. The uncertain geological property, $X$, will be the permeability field [mD]. Hence,~\eqref{eq:optim_prob} gives the maximum storage capacity for [a realization of] the permeability field without exceeding the fracture pressure threshold.

To analyze various risk sharing scenarios, we need to estimate  the distribution of $Y$ by means of an ensemble of $N$ samples,
denoted $\{ \ysamp \}_{n=1}^{N}$, which are functions of the permeability field samples, denoted $\xsamp$. We choose the standard Monte Carlo (MC) method to generate $\ysamp$, due to its simple and non-invasive implementation. A drawback of MC methods is that a large amount of samples is needed for accurate statistical inference. The procedure for generating $\ysamp$ with the MC based algorithm is as follows: (i) generate a realization of the permeability, $\xsamp$ and (ii) solve the optimization problem given in~\eqref{eq:optim_prob}. 
In the following sections we provide some details of steps (i) and (ii), as well as a short description of the storage area used in the numerical test cases. 

\subsection{Geostatistical permeability model}\label{sec:geostat}

To generate realizations of the permeability field, $\{\xsamp \}_{n=1}^{N}$, we follow the procedure standard to the reservoir characterization community and use a geostatistical modeling approach. To this end, we assume that the natural logarithm of the permeability field
is a Gaussian random field. Both the mean log-permeability, $\log\overline{\perm}$, and covariance matrix, $C_{\log\perm}$, are usually defined based on, e.g., interpretations of the storage area and/or analog geological structures. For simplicity, we assume a spatially constant mean field $\overline{\perm}$, and that $C_{\log\perm}$ is generated from an analytical, Gaussian and stationary covariance function given by 
\begin{equation}
    \operatorname{Cov}(h) = \beta\exp(-(h/\gamma)^2),
\label{eq:cov_exp}
\end{equation}
where $h$ is the distance between two points in the storage area, $\beta$ is the variance, and $\gamma$ is the correlation length. To allow for anisotropy $C_{\log\perm}$, i.e., predominant correlations in one of the principal directions, we scale the x-coordinate with a parameter $a_x$, where $a_x=1$ leads to isotropic $C_{\log\perm}$.

A sample of the log-permeability, $\log\perm^{(n)}$, is generated using the Cholesky decomposition method~\cite{Chiles_Delfiner_12}. The parameters used to generate $\log\perm^{(n)}$ in all test cases in~\secref{sec:num-no-comp}-\ref{sec:num-press-comp} are as follows: $\overline{\perm}=1000$ mD, $\beta=0.5$, $\gamma=15.0$ km, and $a_x=0.36$. Note that the permeability field varies in space due to the stochastic variation in the Gaussian field model.

\subsection{Optimization with differential evolution}

To solve the constrained optimization problem ~\eqref{eq:optim_prob}, we used an evolutionary algorithm, namely the  differential evolution (DE) algorithm~\cite{storn1997} with constraint handling from~\cite{lampinen2002}, implemented in the Python package \texttt{scipy}~\cite{2020SciPy-NMeth}.  
DE (and more generally evolutionary methods) does not require derivative computations to generate step vectors in the optimization, but instead rely on iteratively improving candidate vectors to search for the optimal solution. Hence, it is well suited for complex optimization problems where analytical derivatives are not available or costly to compute with sufficient precision.

The overall procedure of the DE algorithm is as follows: (i) create an initial population of $N_P$ candidates; (ii) mutate candidates using a differential variation of two `parent' candidates scaled with a parameter $F$~\cite{qiang2014}; (iii) increase population diversity by accepting vector elements from a mutated candidate with probability $P_{\text{m}}$ to generate trial vectors; (iv) accept trial vectors with higher function values to next generation; (v) repeat (ii) -- (iv) until convergence. The initialization step (i) is done using Latin hypercube sampling~\cite{Eglajs_Audze_77, Mckay_etal_79} to ensure good coverage of the parameter search space, where we set $N_P=20\nctrl$ with $\nctrl$ being the number of control variables. Furthermore, we set $F\sim \mbox{Unif}[0.5,1]$ and $P_{\text{m}}=0.7$ in all test cases.

\subsection{Case study description: Utsira Formation}\label{sec:utsira}
\begin{figure}
    \begin{minipage}{0.4\textwidth}
    \begin{overpic}[width=\textwidth]{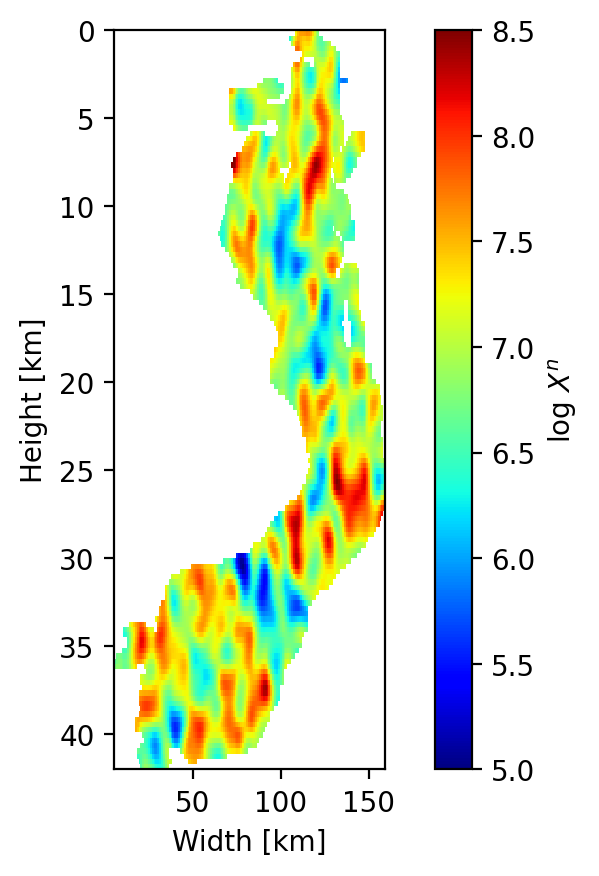}%
    \put(14,90){\large{(a)}}%
    \end{overpic}%
    \end{minipage}%
    \begin{minipage}{0.6\textwidth}%
    \begin{overpic}[width=0.26\textwidth]{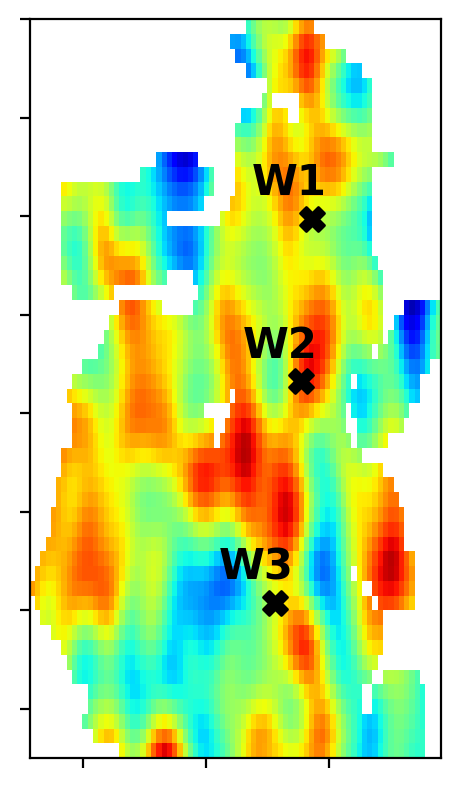}%
    \put(6,85){\large{(b)}}%
    \end{overpic}%
    \begin{overpic}[width=0.2617\textwidth]{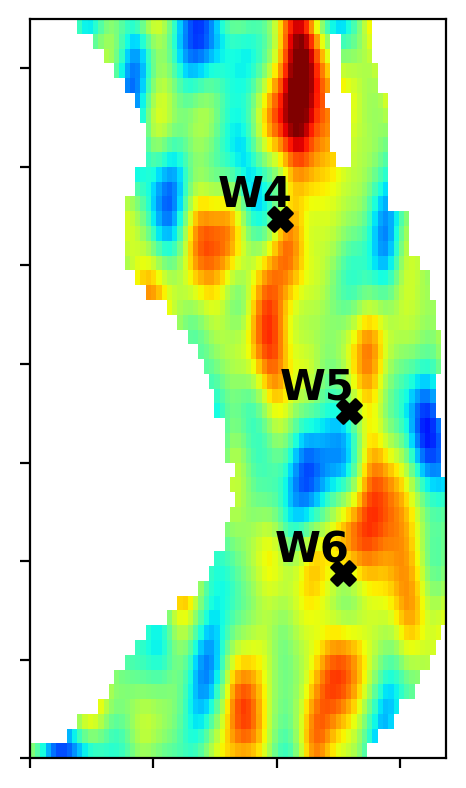}%
    \put(5,85){\large{(c)}}%
    \end{overpic}%
    \begin{overpic}[width=0.495\textwidth]{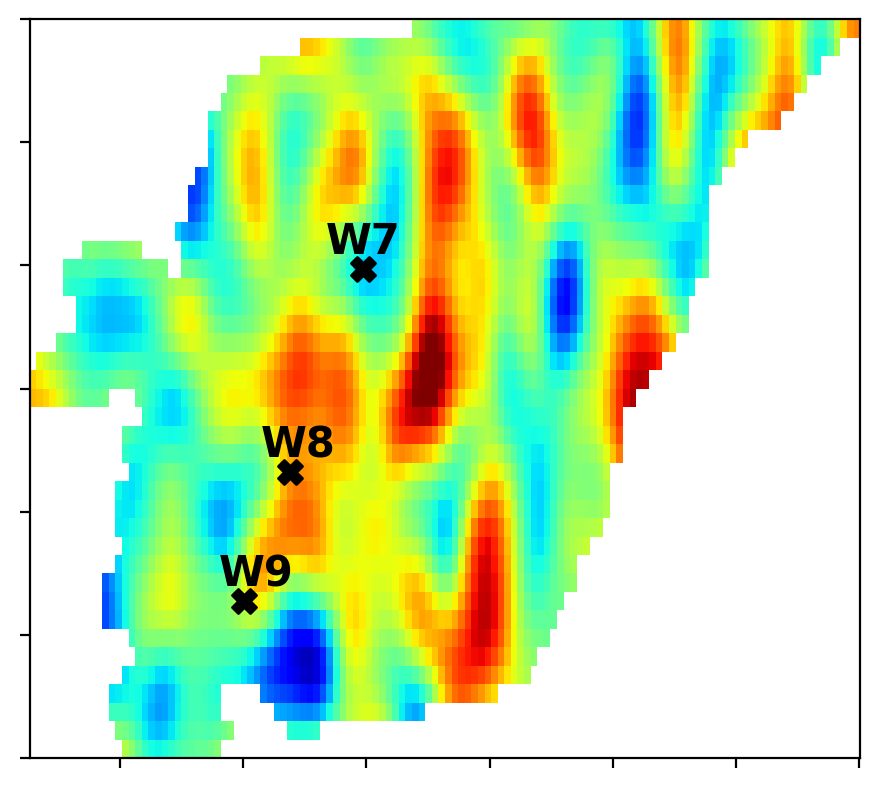}%
    \put(5,75.5){\large{(d)}}%
    \end{overpic}%

    \centering%
    \begin{overpic}[width=0.5\textwidth]{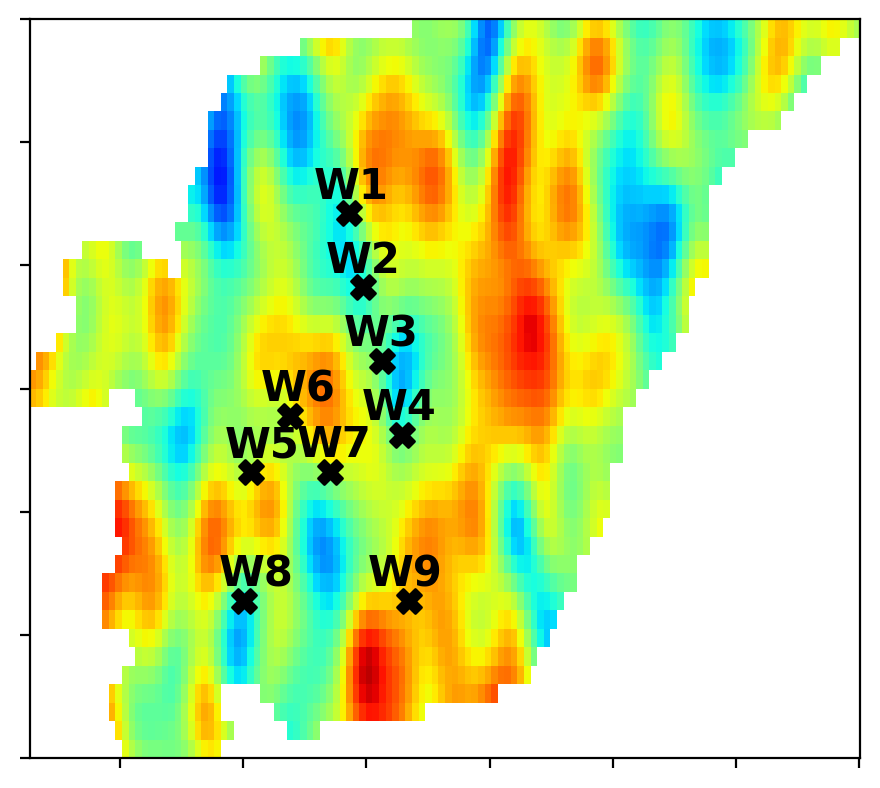}%
    \put(5,75){\large{(e)}}%
    \end{overpic}%
    \end{minipage}
    \caption{(a) An arbitrary permeability realisation for the Utsira formation; well locations for Case I agents (b) $a_1$, (c) $a_2$, and (d) $a_3$; (e) Case II well locations for agents $a_1$ (W1--4), $a_2$ (W5--7), and $a_3$ (W8--9)}
    \label{fig:utsira_new}
\end{figure}

\begin{table}
\begin{center}
\begin{tabular}{lll} \toprule
    {Property} & {Description} & Value \\ \midrule
    \midrule
    Porosity & homogeneous & 0.35 \\
    Pore compressibility & constant & 0.9 GPa$^{-1}$ \\
    Initial pressure & hydrostatic & 100 bar @ 1000 m \\ 
    Water &   end-point value  & 1.0 \\
          &  residual sat. &  0.11 \\
    \co & end-point value  & 0.75 \\
        & residual sat. & 0 \\ 
    \bottomrule
\end{tabular}
\end{center}
\caption{Rock and relative permeability properties used in simulations of the Utsira Formation.}
\label{tab:properties}
\end{table}
The storage area chosen for the test cases in~\secref{sec:num-no-comp} and~\secref{sec:num-press-comp} is the Utsira Formation, located in the Norwegian North Sea. Utsira has been extensively studied in the literature and has been identified as a \co\ storage area with potentially large capacity~\cite{Halland2014}. 
The numerical model is based on data provided by the British Geological survey~\cite{BGS_data}, with relative permeability and rock properties given in~\tableref{tab:properties}. A map of an arbitrary realization of the permeability of Utsira can be seen in~\figref{fig:utsira_new}(a), where we note that the southern part has also been the focus of previous \co\ storage studies, see, e.g.,~\cite{HODNELAND201920}. The 2D spatial domain is a $150\times150$ grid with 8110 active cells (i.e., with non-zero permeability) of size $1070\,\mathrm{m}\times3000\,\mathrm{m}$. The boundaries of the model are closed with a no-flow condition. 
The simulations are initialized fully saturated with water, in hydrostatic equilibrium, and cover injection during 25 years.

Simulations of the \co-water system at Utsira is performed using the Open Porous Media (OPM) Flow simulator~\cite{rasmussen2021open} applying the \texttt{CO2STORE} module~\cite{Sandve2021}. OPM Flow is an open-source, fully-implicit reservoir simulator with industry-standard input and output, and highly scalable parallelization capabilities to handle multi-million cell simulation models. The \texttt{CO2STORE} module uses well established PVT models to accurately calculate fluid properties for \co\ and water, such as dissolution, density and thermal effects.

\section{Case I: Risk sharing by non-competing agents}
\label{sec:num-no-comp}

First consider CO$_2$ injection by competing agents without pressure communication or other interference that would affect the others, which means that their payoffs are well-defined even if they do not collaborate. Furthermore, this is a true stochastic cooperative game, with well-defined and stochastic payoffs for all possible coalition structures. One should note here that unless they do form non-trivial risk sharing schemes, the payoffs will remain identical for all coalition structures. Hence, the dynamics of the games arises from the risk sharing.

We investigate the distribution of total injection into the three groups of wells operated by three agents $a_1$, $a_2$, and $a_3$, respectively, and separated so that there are no constraints from overlapping pressure. Well locations for each agent are shown in~\figref{fig:utsira_new}(b) - (d).
Since the operations have no effect on the other agents, each agent has a well-defined stochastic total injection volume $Y_i^{\text{ind}}$, determined solely by physical limitations. This injection volume $Y_i^{\text{ind}}$ defines a suitable baseline payoff for each of the three agents $i=1,2,3$ when no collaboration occurs, also known as the singleton coalition structure.

\begin{figure}
    \centering 
\includegraphics[width=0.96\textwidth]{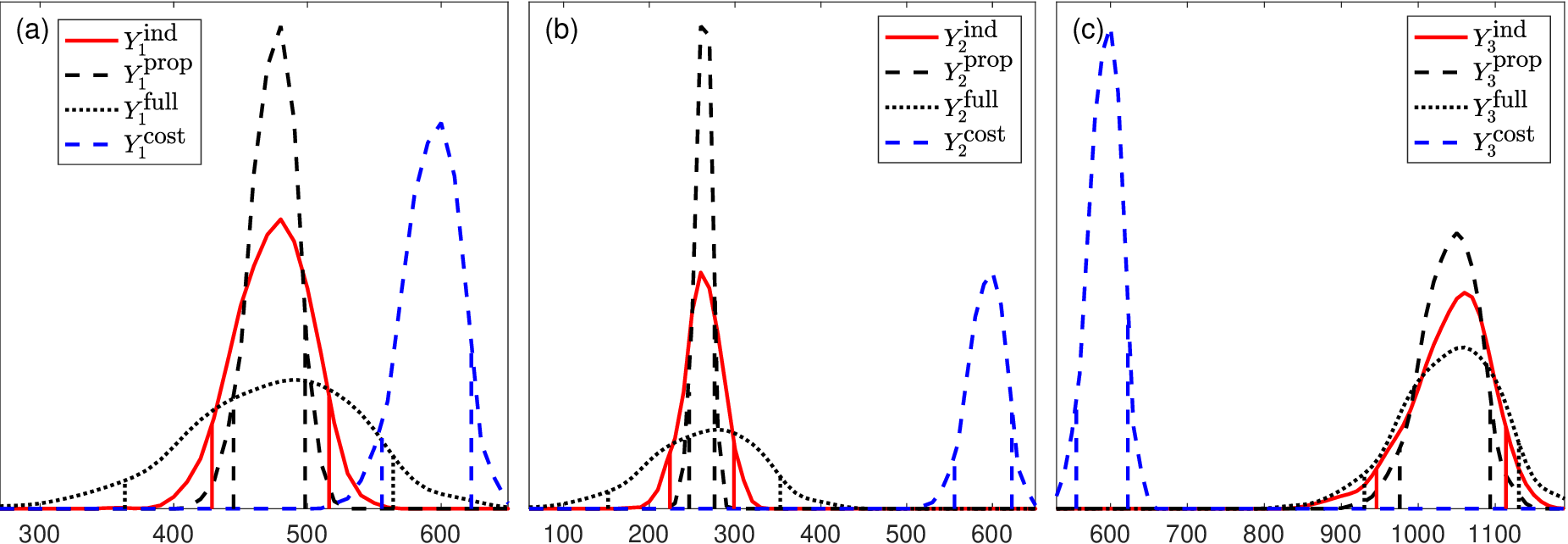}
\caption{Estimated marginal PDFs for the  maximum individual injection $Y_1^{\text{ind}}, Y_2^{\text{ind}}, Y_3^{\text{ind}}$, and different risk-sharing allocations based on fractions of $Y$. Vertical lines denote the 0.1- and 0.9-quantiles, respectively. All injection masses in Mt.
}
\label{fig:1D_PDFs_case_1}
\end{figure}

The individual payoff PDFs resulting from the risk allocations $Y_i^{\text{prop}}$,  $Y_i^{\text{risk}}$, and $Y_i^{\text{cost}}$ are compared to those of the baseline PDFs $Y_i^{\text{ind}}$, as shown in Figure~\ref{fig:1D_PDFs_case_1}. The quantile preference with $\alpha=0.1$ (risk averse) or $\alpha=0.9$ (risk seeking) is assumed henceforth. These quantiles are straightforward to estimate from the order statistics. The exact choice of quantile is not important in a qualitative comparison. However, more extreme quantiles, i.e. $\alpha$ close to 0, would require different methods, e.g. subset simulation for rare-event estimation~\cite{Pettersson_etal_22}.

Consider the two first PDFs of each subfigure of \figref{fig:1D_PDFs_case_1}, i.e., the individual injections $Y_i^{\text{ind}}$ and the contribution-proportional allocations $Y_i^{\text{prop}}$, respectively. For $\alpha=0.1$, all agents prefer the new proportional allocation $Y_i^{\text{prop}}$ over the baseline scenario (Eq.~\eqref{eq:mut_ben_crit} holds), hence there is incentive for all agents to collaborate if they are risk averse. Thanks to the distance between the quantiles (vertical red and dashed black lines) in all three subplots, a subset of agents could reduce their $d_i$ and their $r_i$, while the remaining agents do the opposite, and still satisfy the mutual benefit condition~\eqref{eq:mut_ben_crit}. Such an adjustment leads to a new allocation that is preferable to all agents, again according to the stipulated preference measure. Apparently, the tentative risk allocation $Y_i^{\text{prop}}$ being proportional to $E(Y_i^{\text{ind}})$ was a very good initial guess that led to a feasible (i.e., attractive) solution for risk averse agents. Note that the risk sharing schemes based on concepts such as proportion of investment or physically feasible injection for each agent are meant as intuitive illustrations. This is actually unnecessarily restrictive: what we really could do is to choose any value from the feasible set of $d_i, r_i$, whether that set has any physically/fairness motivated interpretation or not.
The criterion~\eqref{eq:surplus_cond} is satisfied, so the agents can all do better than performing their individual injection according to the baseline scenario. The distribution according to $Y_i^{\text{prop}}$ is just one attractive allocation, but it is not unique.
Note that different choices of quantiles may give quite different results. In particular, for very small $\alpha$ the horizontal distance between the vertical quantile lines will be greater. 

Comparing the same PDFs pairwise, it is clear that risk seeking ($\alpha=0.9$) should not collaborate. For this case, the criterion~\eqref{eq:surplus_cond} is not satisfied, so there are no linear allocations of $Y$ that are attractive for risk seeking agents. Now compare the individual baseline (red) and cost proportional (dashed blue) PDFs in~\figref{fig:1D_PDFs_case_1}. In this case, $a_3$ will not agree to collaborate on investment-proportional terms.


\section{Case II: Risk sharing under pressure competition}
\label{sec:num-press-comp}
Now consider a situation where the three agents’ projects are in hydraulic communication, such that the different operations affect one another through pressure competition. 
Wells 1-4 belong to $a_1$, wells 5-7 to $a_2$, and wells 8-9 to $a_3$, as shown in~\figref{fig:utsira_new}(e).  For this setup, there is typically no unique set of controls that maximize total injection. With pressure competition between the agents, there is no unique solution to the coalition specific allocation based on individual contributions (possible injection) since this is a solution to a multi-objective optimization problem. For more details on related cooperative game models for subsurface resources competition, we refer to~\cite{Pettersson_etal_25}. We can still present the investment proportional allocations of total capacity, as well as different partitions of $Y$ itself, but there are no well-defined PDFs of payoffs proportional to capacity based on physical simulations under geologic uncertainty. This lack of straightforward computation results in imperfect knowledge, which implies a greater reliance on negotiation between the agents to arrive at a satisfactory risk sharing model.  For instance, an agent can claim that its modest contribution is due to being flexible to the operations of others, for which it should be compensated.

In this section, we present an approach that can be used to formulate and evaluate risk sharing models even with imperfect physical knowledge. In particular, we address the challenge of defining the baseline $Y_i^{\text{ind}}$ when agents are are in pressure competition even if they are not collaborating.  Recall that $Y_i^{\text{ind}}$ is used in different risk allocation models presented in Section~\ref{sec:payoff-alloc-models}, and is also used to evaluate the preference relation (Section~\ref{sec:pref-rel}) when ranking different risk sharing alternatives.  It is clear that the individual baselines that could be easily obtained by isolated simulations in Case~I are no longer valid in the pressure competition inherent to Case~II, as each agent’s baseline must now always presume some injection by other agents in the shared resource. In the following sections we describe and analyze the different sources of uncertainty (Section \ref{sec:det_var_planes}), and propose how the lack of baseline can be addressed through (1) a modified approach to  risk allocation  (Section~\ref{sec:utopian-risk-allo}) and (2) the use of belief distributions to address the imperfect knowledge of other competing agents' decisions when no risk sharing occurs (Section~\ref{sec:bel_dist}). We then apply these concepts to the example of three agents in pressure communication in the Utsira Formation (Section~\ref{sec:num_res_bel_Utsira}). 

\subsection{Deterministic payoff variation due to competition}
\label{sec:det_var_planes}

\begin{figure}
    \centering 
\includegraphics[width=0.98\textwidth]{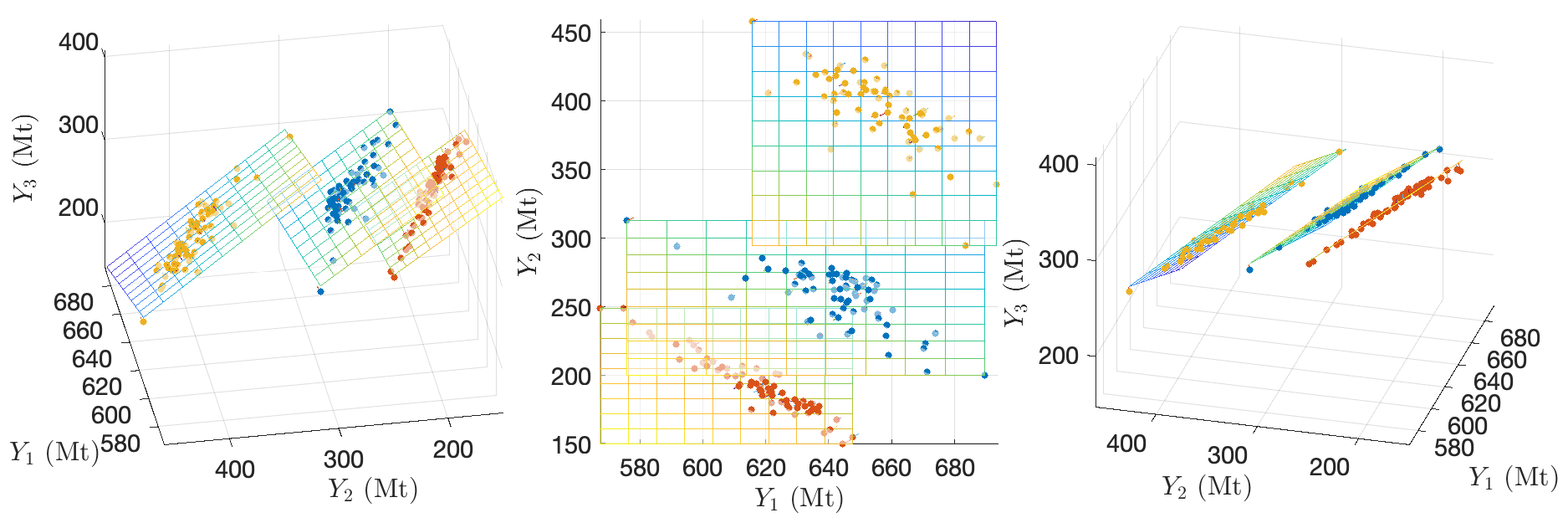}
\caption{Variability in the agents' contributions to total maximum CO$_2$ injection. Points in the same plane corresponds to a fixed permeability field realization. The planes correspond to minimum, maximum, and a randomly sampled permeability. The three planes are shown from different angles for increased visibility.}
\label{fig:Pts_in_plane}
\end{figure}

Performing numerical optimization typically only returns one out of a number of optimal controls. Repeated numerical optimization with stochastic methods may yield a number of different optimal control solutions, but the distribution of these solutions is an effect of the randomness of the numerical approximation, and not a manifestation of some intrinsic geological uncertainty. The situation is illustrated for the Utsira case with communicating wells in Figure~\ref{fig:Pts_in_plane}, where a plane has been fitted to each of a set of points representing numerically computed injection schedules that all lead to near-optimal total injection for a given permeability field. If all points were indeed exactly optimal, the sum of the coordinates anywhere in the plane would equal the true total maximum. The points closest to each plane in Figure~\ref{fig:Pts_in_plane} shows different injection controls for the three agents that all result in nearly the same approximation of the maximum total capacity. The orthogonal distance from a point to the fitted plane indicates the distance from the true maximum capacity, while the spread within the plane indicates the spread in possible combinations of contributions from the three agents, leading to virtually the same total maximum. Each plane corresponds to a different realization of the stochastic permeability field. Hence, with more planes than the three included in the figure, the spread of planes indicates the variability due to the stochastic permeability models. 

Thus, the figure illustrates and decomposes the effect of deterministic injection scheme variability (within plane variation), numerical error in the capacity optimization (distance from point to fitted surface), and the stochastic geological variability (spread between the planes).
While the points identified are indeed feasible nearly optimal solutions, the frequency with which they appear is a direct consequence of the numerical method being used, and not physically motivated stochastic variation. Since the deterministic variability in the injection schedule thus obtained is significant in comparison to the stochastic variability in the sense that that the intra-plane variation is non-negligible compared to the inter-plane variation, direct estimates of the PDFs of $Y_i$ ($i=1,2,3$) are not meaningful.

\subsection{Utopian risk allocation model}
\label{sec:utopian-risk-allo}
The  risk-sharing allocations suggested in Section~\ref{sec:payoff-alloc-models} that seemed natural in the absence of pressure competition, may not be satisfactory in the case of pressure communication. A possible remedy is to define a \textit{utopian} scenario based on physical considerations. Here, the utopian solution is defined as the individually optimal solution for all agents, despite the fact that these solutions cannot be realized simultaneously. The optimal scenario for $a_1$ is the maximized injection into $a_1$'s wells with no injection at all in the wells belonging to $a_2$ and $a_3$. Analogously, the optimal scenario for respectively $a_2$ and $a_3$ is when the remaining agents do not perform injection at all. Denoting the random variable corresponding to the injected CO$_2$ of $a_i$ under these utopian  conditions $Y_i^{\text{u}}$, we clearly have
\[
\sum_i E\mleft(Y_i^{\text{u}}\mright) \geq E(Y),
\]
with equality only if there is no competition between agents. To illustrate this concept, we show the utopian scenario for the three agents operating in pressure competition in the Utsira case in Figure~\ref{fig:tot_1D_PDFs_case_2_ext}(a). It is clear that all utopian injections cannot be simultaneously realized, as they correspond to a total amount of on average twice the physical maximum. 

Despite the fact that agents cannot expect the other agents not to perform injection operations, the relative value of 
$E(Y_i^{u})$ still gives a measure of the injection potential of $a_i$. Also, the distribution of the injection reflects what is physically feasible. 
The utopian injections hence appear suitable for a risk-allocation satisfying Equation~\eqref{eq:payoff_split_Suijs}, and being applicable to the case of pressure communication, unlike some other allocations proposed in Section~\ref{sec:payoff-alloc-models}. Hence, we propose setting
\[
d_i = \frac{E\mleft(Y_i^{\text{u}}\mright)}{\sum_k E\mleft(Y_k^{\text{u}}\mright) } E(Y), \quad r_i = \frac{E\mleft(Y_i^{\text{u}}\mright)}{\sum_k E\mleft(Y_k^{\text{u}}\mright)},
\]
leading to
\[
Y_i^{\text{u, prop}} \equiv \frac{E(Y_i^{\text{u}})}{\sum_k E\mleft(Y_k^{\text{u}}\mright) }Y.
\]
Again, if no resource competition occurs, $Y_i^{\text{u}}=Y_i^{\text{ind}}$, and the above risk sharing scheme is equivalent to~\eqref{eq:risk-sharing-phys-prop}.

\subsection{Belief distributions for non-collaborative injection}
\label{sec:bel_dist}
As motivated above, there is no unique baseline scenario with actual injection that defines the payoffs of individual agents that choose not to collaborate (or any other subcoalition of all agents for that matter). If one can construct probability distributions for this situation, they can be compared to the collaborative case using a preference relation, so that one can make an informed decision about whether to collaborate.
This section addresses how a baseline can be defined in the case of  incomplete information.

To motivate the concept of belief distributions, let us consider an analogy. Assume we want to throw a standard die of the kind used in most board games. What is the outcome? Well, it depends on the initial position of the die in the hand of the person throwing it, the force used when exiting the hand, the surface on which it falls, and several other factors. It probably appears to most readers that even a very elaborate physical model may fail to deliver the correct result. In contrast, it seems to be almost universally accepted that unless somebody has tampered with the die, a stochastic model assigning probability 1/6 to each of the six faces, is an adequate model for the situation of interest. Indeed, such a model is also simple and easy to evaluate, especially compared to the first option. What this example is meant to illustrate is that situations that are in reality completely deterministic (any instance of the throwing of a die could in principle be exactly characterized by deterministic physics) can in some cases be more appropriately be described by a stochastic model. This is in particular true if the stochastic model is simple (or much simpler), and provides a range of outcomes that both accounts for errors any deterministic model may exhibit, as well as the full range of possible outcomes.

In the case of pressure communication, the injection of one agent typically depends on the actions of other agents. While the actual outcome ultimately depends deterministically on the actions and reactions regarding injection of all agents involved, by analogy if the die example above, it may be useful to form a \textit{belief distribution} by means of subjective probability that defines a stochastic model for the range of different outcomes when no collaboration occurs. 

In the following, we will outline the properties that belief distributions should possess, how they can be constructed, and ultimately come up with two concrete suggestions for belief distributions. In practice, these example distributions are of course only useful if practitioners agree with the assumptions made so that they do reflect actual beliefs. We emphasize that they could and should be replaced with the practitioners' own belief distributions whenever they do not align with the proposed distributions.

Before we proceed to propose belief distribution models, we make a brief note on the close relationship between stochastic preference relations, subjective probability, and belief distributions, as investigated in a series of works following~\cite{Savage_54}. 
According to a series of related and widely used definitions of subjective probability and preferences, the subjective probability of an agent faced with an uncertain situation can be directly inferred from the agent's different preferences with respect to that situation. For instance, if the agent prefers action $A$ over action $B$ in a game, it can be inferred that the subjective probability of winning is higher with action $A$ compared to $B$. 
This is in contrast to the use of subjective probability or belief distribution used in the current work, which is not defined in terms of preferences. Any agent will prefer a larger injection volume over a smaller injection volume, but the agent may not assign a larger probability to that being the case unless there is some compelling evidence that this would indeed happen.

The term belief has a technical definition in Dempster-Shafer theory and a bit simplified denotes the (perceived) probability that supports a certain outcome. This is in contrast to the term plausibility, which is one minus the probability of events that imply that the outcome in question does not occur. In that setting, belief and plausibility are respectively lower and upper bounds on probability. In contrast, in the current work the belief distribution is simply the best estimate of the probability distribution of what an agent believes will happen. This is in agreement with the ideas of Aumann models of incomplete information described in detail in e.g.~\cite{Maschler_etal_20}.

Among the outcomes in terms of non-collaborative injection, any given agent will naturally prefer certain scenarios over others, i.e., more injection over less injection of CO$_2$. We will however assume that the fact that an agent would prefer that the others leave some more pressure space unaffected has no effect as those other agents would prefer to use that pressure space themselves. Furthermore, we will assume that all agents share a joint belief model. The agents' individual prospects may vary, and their actions too, but they all agree on what is possible and what the effects would be. This leads to a consistent albeit subjective shared model for the relevant distributions of injection under full competition that can be compared to the distributions of injection under collaboration, by means of the agents' stochastic preference relations.

To be useful, a belief distribution should reflect what can and cannot happen due to physical constraints, and include whatever other information available. It should be informative in the sense that it does not lead to an overtly wide distribution, but all assumptions narrowing the distributions should of course be clearly justified in order to not create a false sense of safety. The belief distribution for non-collaborative injection is almost by definition wider than the jointly optimized injection distribution under the grand coalition. The former includes both the geological uncertainty of the latter, and assumed distributions for the actions of external agents. Hence, it seems natural that in general risk averse agents will prefer collaboration, and risk seeking agents will prefer non-collaboration. There may however be important exceptions to this rule-of-thumb, and it may also be useful to quantify the choices and make an informed rigorous decision based on the quantitative model.

According to the maximum entropy principle, for lack of more precise information, one should always choose the entropy maximizing (i.e., the least informative) probability distribution from the class of distributions that fit the available prior information. In this way, we avoid assumptions that are not supported by data or other sources of reliable information. For instance, we do not assume Gaussian distributions for the sheer reason of convenience, but only if there is compelling reason to do so. If a quantity of interest is restricted to an interval, the entropy maximizing distribution is the uniform distribution on that interval, implying that all sub-intervals of equal length are considered equally likely. This is a reasonable assumption since there is no evidence for certain intervals being more likely than others. 
 
 With two or more agents, a possible model is to start from the assumption that the payoff of one agent is independent of the payoff of another agent. The range of possible payoffs then forms a rectangle in the case of two agents, and a hyper-rectangle in the case of multiple agents. Since it not a realistic feature that e.g. all agents inject as if no one else is present on site, additional constraints can be imposed to account for such features, resulting in a modified shape of the originally hyper-rectangular domain. Alternatively, to account for the fact that increased injection by one agent typically needs to be compensated by reduced injection by another agent, a simplex-shaped domain that directly imposes such constraints may be suitable. For a quantity of interest on a simplex, the entropy maximizing distribution is the so-called flat Dirichlet distribution, i.e., uniform density on the simplex. Uniform probability densities on either a modified hyperrectangle or on a simplex will be the basis for the belief distributions to be proposed in this work.

Next, we discuss imposition of stochastic upper and lower bounds on the uncertain injection intervals. 
For a given realization of the uncertain geological parameters, an agent can at most inject its theoretical maximum that assumes that all other agents comply and that the agent in question is only limited by the physical constraints. This is then a reasonable upper bound on what the agent can inject. Even in the worst case where other agents severely limits what an agent can inject, there may be room for some injection, described by a worst-case injection random variable $Y_i^{\text{wc}}$ that can be estimated from simulation data. For lack of actual estimates of this capacity, subject to constraints from other agents maximizing their own revenue, a conservative lower bound is simply to assume zero injection, i.e., $Y_i^{\text{wc}}\equiv 0$. Without further information we may then assume that all injection scenarios between the worst-case and utopian scenario are equally likely, and set
\[
Y_i^{\text{bel}}
=
Y_i^{\text{wc}} + (Y_i^{\text{u}}-Y_i^{\text{wc}})U_i
=
(1-U_i) Y_i^{\text{wc}} + U_i Y_i^{\text{u}}, \quad U_i \sim \mbox{Unif}[0, 1].
\]
To be realistic, a belief model $Y_i^{\text{bel}}$ should not allow payoffs exceeding the possible maximum, i.e., they should satisfy
\[
\sum_{j} Y_j^{\text{bel}} \leq Y.
\]
Hence, we should identify (dependent) random variables $U_j$ ($j=1,\dots, \Na$) so that the above inequality holds.  

\subsubsection{Belief distribution on simplices with uncertain vertices}
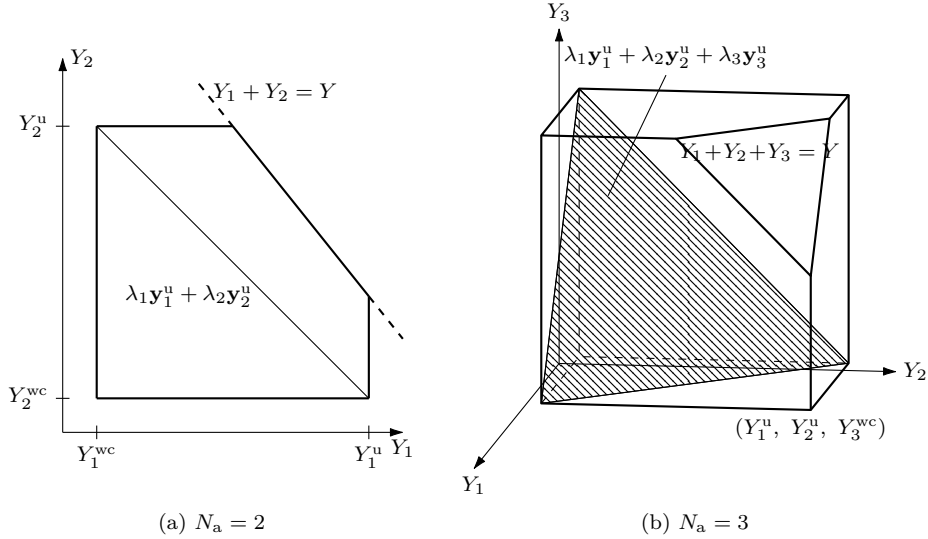
\begin{figure}[ht]
\tikzset{>={Latex[round, length=2mm,width=1.2mm]}}
\centering  
\subfloat[$\Na=2$]{ 
{\scalefont{0.8}
\begin{tikzpicture}[scale=0.90]

        \draw[thick, black] (1,3) -- (1,7);
        \draw[thick, black] (1,3) -- (5,3);
        
        \draw[thick, black] (1,7) -- (3,7);
        \draw[thick, black] (5,3) -- (5,4.5);
        
        \draw[thick, black]  (3,7) -- (5,4.5);
        \draw[thick, black, dashed] (2.5,7.625) -- (3,7);
        \draw[thick, black, dashed]  (5,4.5) -- (5.5,3.875);
        
        \draw (2.6,7.5) node[right]{$Y_1+Y_2 = Y$};
        
         \draw[thin, black] (1,7) -- (5,3);

 	\draw (3.4,4.5) node[left]{$\lambda_1 \mathbf{y}_1^{\text{u}} + \lambda_2 \mathbf{y}_2^{\text{u}}$};
  
         \draw[->] (0.5,2.5) -- (5.5,2.5) node[below] {$Y_1$}; 
         \draw[->] (0.5,2.5) -- (0.5,8) node[right] {$Y_2$};
         
         \draw[thin, black] (1,2.4) -- (1,2.6); 
         \draw[thin, black] (1,2.4) node[below]{$Y_1^{\text{wc}}$};
         
         \draw[thin, black] (5,2.4) -- (5,2.6); 
         \draw[thin, black] (5,2.4) node[below]{$Y_1^{\text{u}}$};
         
         \draw[thin, black] (0.4,3) -- (0.6,3); 
         \draw[thin, black] (0.4,3) node[left]{$Y_2^{\text{wc}}$};
         
         \draw[thin, black] (0.4,7) -- (0.6,7); 
         \draw[thin, black] (0.4,7) node[left]{$Y_2^{\text{u}}$};
         
         \draw[color=white] (1,1.5) -- (4,2);
        
\end{tikzpicture}
}
}
%
%
\subfloat[$\Na=3$]{ 
{\scalefont{0.8}
    \begin{tikzpicture}[scale=0.90, 3d view={98}{10},line join=round,
        declare function={a=4;b=2;}]
        \draw[style=dashed, color=black] (a,0,-a) -- (0,0,-a)-- (0,a,-a);
        \draw[style=dashed, color=black] (0,0,0) -- (0,0,-a); 
        \draw[thick,black] (a,0,0) -- (a,0,-a) -- (a,a,-a) -- (a,a,-b);
        \draw[thick,black] (a,a-b,0) -- (a,0,0) -- (0,0,0) -- (0,a,0) -- (a-b,a,0);
        \draw[thick,black] (0,a,0) -- (0,a,-a) -- (a,a,-a);
        
        \draw[thick] (a,a,-b) -- (a-b,a,0) -- (a,a-b,0) -- cycle;
        
        \draw[thin, color=black] (a,0,-a) -- (0,0,0);
        \draw[thin, color=black] (0,0,0) -- (0,a,-a);
        \draw[thin, color=black] (0,a,-a) -- (a,0,-a);      
         
         \draw[thin,pattern=north west lines] (a,0,-a) -- (0,0,0) -- (0,a,-a) -- cycle;
         
         \draw[->] (-5,-1,-5) -- (4,-1,-5) node[below] {$Y_1$}; 
         \draw[->] (-5,-1,-5) -- (-5,4,-5) node[right] {$Y_2$};
         \draw[->] (-5,-1,-5) -- (-5,-1,0) node[above] {$Y_3$};         
        
         \draw (a,a,-a) node[below] {$(Y_1^{\textup{u}},\ Y_2^{\textup{u}}, \ Y_3^{\textup{wc}} )$};

         \draw[thin] (0.5,0.5,-1.5) -- (1.5,1.5,0.5); 

        \draw (1.5,1.5,0.5) node[above] {$\lambda_1 \mathbf{y}_1^{\text{u}} + \lambda_2 \mathbf{y}_2^{\text{u}} + \lambda_3 \mathbf{y}_3^{\text{u}}$};

         \draw (3,3.1,-0.6) node[above]{$Y_1\!+\!Y_2\!+\!Y_3=Y$};

    \end{tikzpicture}
    }
    }
    
   \caption{Schematic depiction of the ranges of the belief distributions for $\Na=2$ (a), and $\Na=3$ (b). The boundaries of the distribution ranges are themselves stochastic.}
   \label{fig:schematic_belief_dist}
   \end{figure}

As a first proposal for a belief distribution, we will assume that each agent will get a relative fraction of their injection potential, denoted by $0 \leq U_i \leq 1$ so that $\sum_i U_i = 1$. Here, $U_i=0$ implies worst-case injection, and $U_i=1$ implies utopian injection for $a_i$.
For instance, with two agents and $U_1 = 1 - U_2 \equiv U$ being uniform on [0, 1]:
\[
\begin{aligned}
Y_1^{\text{bel}}
= 
(1-U) Y_1^{\text{wc}} + U Y_1^{\text{u}}
\\
Y_2^{\text{bel}}
=
U Y_2^{\text{wc}} + (1-U) Y_2^{\text{u}}
\end{aligned}
\]
In particular, if $a_1$ increases its injections by an amount $\Delta U (Y_1^{\text{u}}-Y_1^{\text{wc}})$, then $a_2$ must reduce its injections by the amount $\Delta U (Y_2^{\text{u}}-Y_2^{\text{wc}})$. Note that the increase is typically not compensated by an equal decrease by the other agent.
This can be generalized to any number of agents $\Na$ using barycentric coordinates. Let $\mathbf{y}_{i}$ denote the random point in $\Na$-dimensional space with Cartesian coordinates
\[
[\mathbf{y}_{i}]_{j} = 
\left\{
\begin{array}{ll}
Y_{i}^{\textup{u}} & \mbox{ if } j = i,\\
Y_{i}^{\textup{wc}} & \mbox{ if } j \neq i
\end{array}
\right..
\]
With $\mathbf{y}=(\mathbf{y}_{1},\dots, \mathbf{y}_{\Na})$ we then define the entropy-maximizing uniform distribution on the $(\Na-1)$-simplex $S(\mathbf{y})$ as follows:
\[
\mathbf{Y}^{\textup{bel}(S)} \equiv \left(Y_1^{\textup{bel}(S)},\dots, Y_{\Na}^{\textup{bel}(S)}\right) \sim \mbox{Unif}(S(\mathbf{y})), \quad  S(\mathbf{y}) = \sum_{i=1}^{\Na} \lambda_i \mathbf{y}_{i}^{\textup{u}}, \quad \lambda_i \geq 0 \forall i=1,\dots, \Na,\quad \sum_{i=1}^{\Na} \lambda_i = 1.
\]
The belief distribution for the total injection is the sum of the beliefs about the individual contributions: $Y^{\textup{bel}(S)} = \sum_{i=1}^{\Na}Y_i^{\textup{bel}(S)}$.
 The distribution of $\mbox{Unif}(S(\mathbf{y}))$ is schematically depicted in Figure~\ref{fig:schematic_belief_dist}, where the resulting simplex is a line with uncertain end points for $\Na=2$ (a), and a triangle with uncertain vertices for $\Na=3$ (b). Since the simplex $S(\mathbf{y})$ itself is stochastic, the distribution of $\mathbf{Y}^{\textup{bel}(S)}$ is highly non-uniform. The model $\mathbf{Y}^{\textup{bel}(S)}$ is relatively pessimistic in the sense that the total injected volume is at most equal to the maximum sum of the utopian injection of one agent and the worst-case injections of the others. The global maximum $Y$ obtained under coordination can never be realized in this stochastic competition model. This may or may not be a limitation depending on the agents' actual beliefs.

\subsubsection{Belief distribution on polytopes with uncertain boundaries}
As an alternative model, where the global maximum $Y$ can indeed be attained, despite the lack of collaboration, we consider injections that are confined by the following inequalities:
\begin{equation}
\label{eq:bel_polygon}
\begin{aligned}
Y_i^{\text{wc}} \leq Y_i & \leq Y_i^{\text{u}},\quad i=1,\dots, \Na,\\
\sum_i Y_i & \leq Y.
\end{aligned}
\end{equation}
This defines a polytope $P$ in $\Na$-dimensional space: the first line of Eq.~\eqref{eq:bel_polygon} defines a hyperrectangle, the second line defines a plane that cuts off a corner of the hyperrectangle. We know that no safe and reasoanble injection will occur outside that polytope. There may be points within the polytope that cannot be reached for practical reasons, and there may be points that in practice are much more likely to be realized than others. However, for lack of actual information about that, it makes sense to assume that since anything within the polytope can happen (or, rather: nothing outside the polytope can happen), we assign equal probability to all events (including scenarios that cannot really happen, but instead making sure nothing is missed). This is a conservative but still simulation-informed model for injection without collaboration, and under pressure competition. Hence, we propose:
\[
\mathbf{Y}^{\textup{bel}(P)} \equiv \left(Y_1^{\textup{bel}(P)},\dots, Y_{\Na}^{\textup{bel}(P)}\right) \sim \mbox{Unif}(P), \quad \mbox{and} \quad Y^{\text{bel}(P)} \equiv \sum_{i=1}^{\Na}Y_i^{\text{bel}(P)}.
\]
The random variables $\mathbf{Y}^{\textup{bel}(P)}$ thus defined are dependent, reflecting the effects of the actions of one agent on the others. To estimate their distributions, one can use rejection sampling, where one discards all hyperrectangle samples that do not satisfy the total maximum constraint. 

\subsection{Numerical results: belief distributions for Utsira Formation}
\label{sec:num_res_bel_Utsira}
\begin{figure}[ht]
    \centering 
\includegraphics[width=0.98\textwidth]{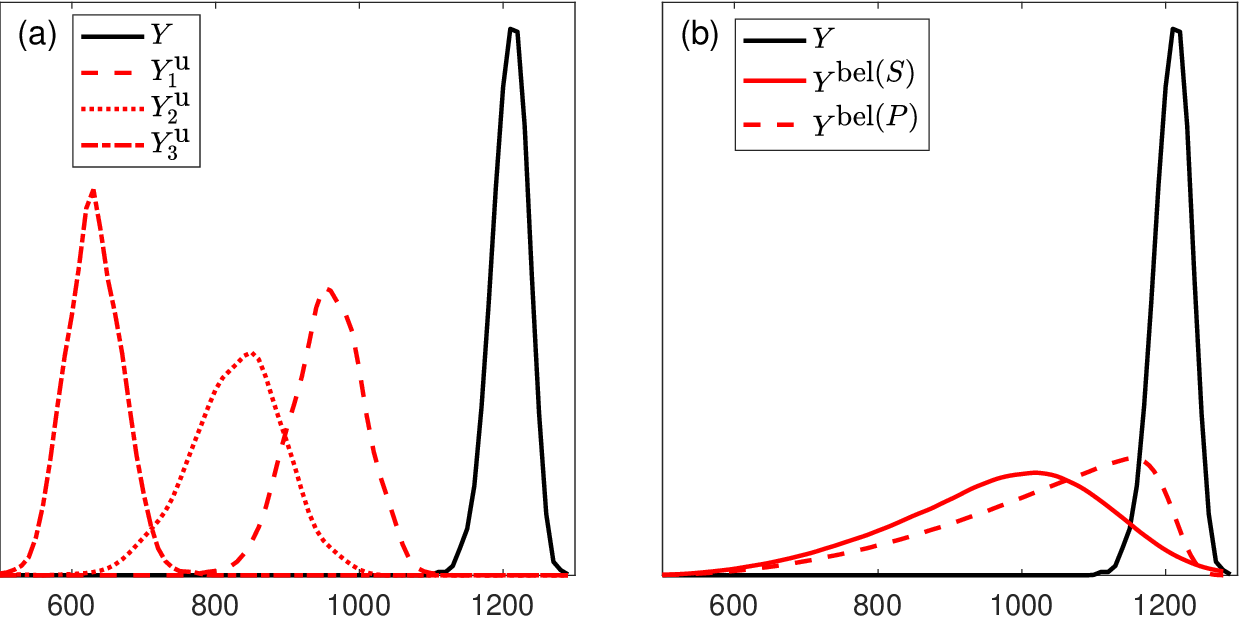}
\caption{PDFs for the total maximum injection $Y$, and utopian injections of the agents one-by-one (a). Maximum and belief distributions for total injections (b).}
\label{fig:tot_1D_PDFs_case_2_ext}
\end{figure}

Next we compute the utopian distributions $Y_i^{\textup{u}}$ for the three agents in the competitive Utsira case. They will be used both to find risk allocations proportional to the utopian injections in the collaborative setting, and to form belief distributions in the competitive setting, for lack of any simple baseline scenarios.
Figure~\ref{fig:tot_1D_PDFs_case_2_ext} shows the PDFs for the  utopian injections of all agents, and the PDF of the total maximum injection. It is clear that all utopian injections cannot be simultaneously realized, as they correspond to a total amount of on average twice the physical maximum. The two belief distributions assume a worst case scenario of 20 \% of the utopian injection rates for all agents, and the stochastic upper bound given by their respective utopian injections. While the upper bounds can be rationally justified, the 20 \% lower bounds are somewhat arbitrary, and should be replaced by distributions from worst-case simulations. As expected, the resulting belief distributions have larger variance and smaller expectation than those of $Y$. The mode of $Y^{\textup{bel}(P)}$ is not far from that of $Y$.

\begin{figure}[ht]
    \centering 
\includegraphics[width=0.98\textwidth]{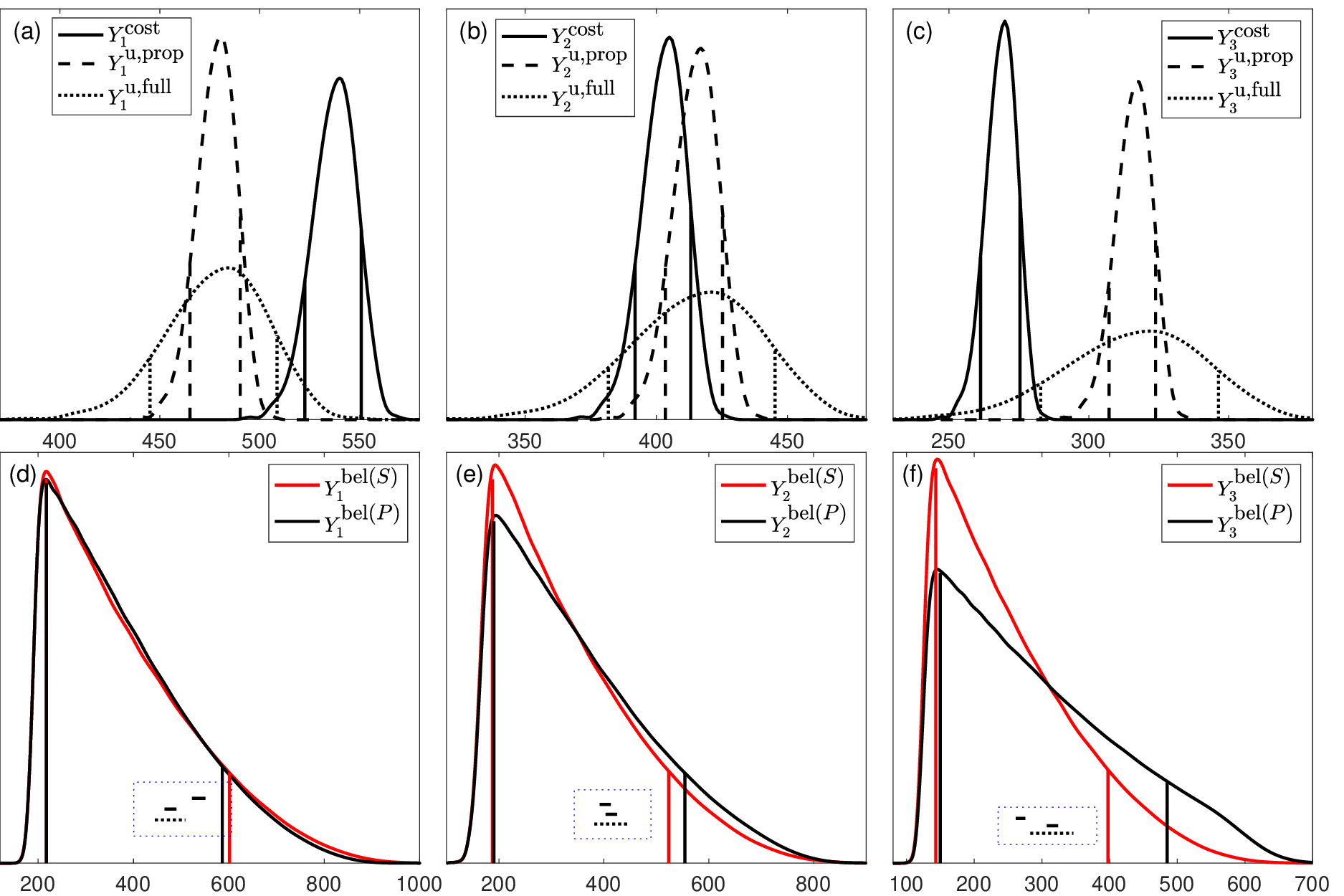}
\caption{Estimated marginal PDFs for the allocations of payoffs as mean-weighted fractions of total injection $Y$ (upper row), and marginal belief distributions for payoffs under pressure competition (lower row). Vertical lines denote respectively 0.1-quantiles, and 0.9-quantiles. The 0.1-0.9 ranges from the upper row plots are indicated by horizontal lines in lower row plots.}
\label{fig:1D_PDFs_case_2_ext}
\end{figure}

The various suggested allocations of the total payoff are shown in  Figure~\ref{fig:1D_PDFs_case_2_ext}~(a,b,c), together with their 0.1 and 0.9 quantiles. In addition to the previously proposed risk allocations, we show the PDFs with mean values proportional to the relative utopian injections, but with the full risk, denoted $Y_i^{\text{u,full}}$. The lower row shows the corresponding belief distributions for each agent. Thus, the distributions of collaborative risk sharing in the upper row can be compared directly to the belief distributions of competitive individual injections in the lower row. Overall, due to the large variability of the belief distributions, a risk-averse agent would want to collaborate for risk sharing, and a risk seeking agent would prefer not to collaborate with the hope of maximizing its own injection. 

\section{Discussion and outlook}
\label{sec:discussion}
Next we discuss some selected topics in more detail, with focus both on limitations, how to overcome them, and possible extensions.

\subsection{Extensions with respect to related game models}
In previous work on deterministic cooperative games for CO$_2$ injection, we mapped the full Pareto fronts of the solutions of multi-objective optimization problems describing injection schemes for different coalition structures under pressure competition~\cite{Pettersson_etal_24, Pettersson_etal_25}. In the current work,  
the focus is on whether agents should join the grand coalition for full collaboration, or not collaborate. The cases of collaboration between some but not all agents have not been explicitly investigated, but is a straightforward extension of the proposed methodology. Without pressure competition, the existing data can be used without modification to investigate the remaining coalition structures. In the case of pressure competition, the utopian and worst-case distributions should be estimated from simulation.

In the case of full collaboration, the uncertain total value is given by a single unique solution, namely the maximum storage capacity is attained, in the game theory literature referred to as social welfare maximization. However, different individual injection schedules can lead to the same maximum total injected CO$_2$ volumes. We did not cover possibly Pareto optimal solutions that are not welfare-maximizing, e.g., when one agent improves its own injection at the expense of the total amount stored, as done in~\cite{Pettersson_etal_25}, or the situation where different agents have different information, leading to possibly non-Pareto optimal solutions~\cite{Chen_Hosseini_25}. Both these situations are however amenable to the stochastic payoff framework.

\subsection{Risk allocation strategies and preference measures}
In the risk allocation step (insurance stage), the uncertain total value is distributed to the individual agents. We have suggested a few allocation schemes proportional to anticipated individual injection without competition or investment costs, typically assuming the same proportionality for the risk and the deterministic part of the payoffs. This is however not necessary. Any risk allocation is possible as long as the sum of all contributions equal the total payoff.

We have assumed quantile preferences for all agents in the analysis and numerical experiments. If other relations reflecting the agents' true preferences can be identified, they should of course be applied instead. Even if they do not easily lend themselves to theoretical analysis of arbitrary random variables, one could still perform a mapping by applying them for a range of allocations to the numerically estimated stochastic payoffs (total CO$_2$ injections), and approximating the indifference curves for subsequent analysis.

In some cases, preferences have properties that admit a so-called certainty equivalent, i.e., a deterministic monetary value so that the agent is indifferent between receiving the stochastic payoff or the certainty equivalent~\cite{Suijs_Borm_99b}. For the quantile preference $\succeq_{\alpha}$, the certainty equivalent of a random variable $A$ is equal to $q_{\alpha}^{A}$.
Note that an agent is not necessarily very happy to receive its deterministic equivalent: it is just equally (un)happy to receive the deterministic equivalent as the stochastic payoff. A risk averse agent (with a smaller $\alpha$) will accept a lower deterministic payoff as equivalent of a random payoff, but it surely does not mean that the agent for that matter is willing to accept either the payoff or its deterministic equivalent.

The individual payoff allocations given by the random variables $Y_i^{\text{ind}}$ can have very different and even correlated distributions,
a situation that  can be handled in the proposed framework without any additional challenges. For the results presented in Case I, the distribution of all $Y_i^{\text{ind}}$ appear qualitatively similar, and by the construction of the stochastic geological model, there is  no correlation between them. In situations with pressure communication, the proposed belief distributions exhibit strong correlations to reflect the fact that one agent's increased operations often requires reduced injection by competing agents.

We have considered risk allocations that are linear in the total payoff, so that all agents receive a payoff being a scaling of the same random variable. In principle, payoffs described by very different random  variables can be assigned to the agents, as long as the sum has the same distribution as that of the total injection. That could be attractive for cases where, e.g., one agent's baseline payoff has a multimodal distribution, and one wishes  not to share that risk feature with the other agents.

\subsection{The role of a stochastic baseline scenario and pressure competition}
In the current work, we have emphasized the role of a stochastic baseline scenario corresponding to no collaboration (i.e., singleton coalitions) which is the reference against which an action leading to a different random situation should be compared. Such a baseline scenario has not been assumed in the existing literature on cooperative games with stochastic payoffs, where all payoffs are described by well-defined random variables. Indeed, if the random payoffs are clearly defined for all possible coalitions of agents, it may be possible to search the optimal allocation, i.e., the one that cannot be improved by any other allocation. Since the individual contributions of the well operators optimally injecting CO$_2$ are not unique in the case of pressure interaction  -- more than one injection schedule leads to (nearly) the same total optimum -- there is no natural unique way to determine individual agents' payoffs, and no baseline scenario. While never strictly necessary to find the optimal payoff allocations, establishing a baseline scenario is a practical manner to make sure all coalitions' payoffs are well defined, and simplifies the potentially very conceptually difficult problem of finding the optimal allocation, to finding an allocation that is better than any individual agent can attain on its own.

In the case of no pressure-space competition, the corresponding deterministic game would be trivial since the injection volumes remain the same whether the agents collaborate or not. Hence, all incentive to collaborate as encoded via the preference measure is completely due to the uncertainty in the payoffs. This does however not need to be the case. One could think of situations where the expected payoffs vary between the coalition structures, and where the expectation preference (that would coincide with the corresponding deterministic game) yield a result different from, e.g., the quantile preference.

In the case of pressure competition between injecting agents, we have illustrated the variability due to i) the stochastic geological model; ii) the non-uniqueness in injection schedules that all result in the  optimal total injection volume; and iii)  error in the numerical approximation of the optimal injection schedules. By depicting the injection schedules as points in parallel fitted planes, the three types of variability can be visualized as, respectively the distribution of the planes (i), the spread within the planes (ii), and the distance between the points and their planes (iii). This kind of visualization can be used among other things to check that the numerical optimization errors are not too large compared to other variability not due to discretization errors. Conversely, it can be used to not set optimization tolerances overtly strict, otherwise leading to unnecessarily high numerical cost.   

\subsection{Risk and losses}
We have employed a definition of risk that is roughly equivalent to the stochastic variability in injection volumes. In many situations, most agents would want to avoid leakage of CO$_2$ from the storage site, and it would then be natural to include this in the definition of risk. While an exact definition of risk including leakage is beyond the scope of this work, a suggestion was proposed in Section~\ref{sec:leakage}, and some general remarks can be made. The stochastic payoffs should still be described by random variables, so all aspects of risk need to be condensed to a single random variable. If a fixed cost can be assigned to every possible leakage scenario (e.g., a unit cost per leaked amount of CO$_2$), the payoffs could simply be the sums of the value of the amount of CO$_2$ injected and the penalty for leakage in the cases that happens, everything estimated from the same numerical simulations.

\subsection{Practical implications for decision-making}
As stated in the Introduction, the objective of this study was to gain insight into how risk sharing agreements can be analyzed in the {\co} storage context. We showed that is possible to adapt mechanisms stemming from cooperative game theory that are commonly employed in the insurance industry, and to carry out the analysis in a computationally tractable manner given the available information.  We have found that in the absence pressure competition, the classic theory is directly applicable and requires no additional calculations beyond each agent's own baseline PDF that can be calculated from the local geological uncertainty. The case of pressure competition adds a level of interdependence between the agents that renders the available information incomplete. Namely, the number of "what-if" scenarios between collaborating or not collaborating grows to an intractable level and complicates a straight-forward decision-making process. However, we showed that belief distributions provide a remedy, although imperfect and fuzzy, and can be useful if constructed with care. 

In this paper, we present risk sharing as a way to divide an uncertain shared value (total {\co} stored) among the agents, noting that each agent's earned revenue no longer has any direct relation to that the physical quantity of {\co} stored in their own wells. Thus, collaboration requires the ability to flexibly move {\co} between different projects as the physical system dictates. In practice, collaboration on subsurface risk needs to be facilitated by collaboration on above-ground {\co} transport from the emitters to the projects. Clearly, a shared transport infrastructure is easier to implement if the projects are co-located in the same region. An example of shared infrastructure is between the Dutch offshore projects Porthos and Aramis, where successful implementation can lay the groundwork for transferring {\co} between projects as a risk mitigation measure, noting that Porthos and Aramis are not in pressure communication. Their agreement is not publicly available information, but it points to the awareness and willingness of the industry to collaborate on risk, and therefore the relevance of this study for providing insight into more complicated subsurface scenarios. 

It should be emphasized that the CCS industry is still in its nascent stages, and analysis of risk sharing with neighboring projects may not be an immediate concern for most storage operators. Currently, the risk of lost commercial value is likely incorporated into each individual company's overall investment portfolio that will be dominated by oil and gas for some time still. Nonetheless, the main findings of this paper can provide useful guidance to storage operators as the industry matures. Implementation of this paper's concepts by industry will be of course highly dependent on each companies' inherent decision-making philosophy and associated workflows.

\section{Conclusions}
\label{sec:conclusions}

A decision making and risk sharing framework for CO$_2$ injection under geological uncertainty with multiple injecting agents based on stochastic cooperative game theory has been introduced and demonstrated on a basin-scale site, the Utsira Formation in the North Sea.

In situations of no pressure competition or other interference between operations, risk averse agents should typically collaborate to share risk under the assumed quantile preference measure, unless their injection potential is relatively insensitive to the uncertainty in geological properties. Risk seeking agents should only enter collaboration with a resource allocation that increases their chances of large gains, and that typically requires other agents to take a risk averse attitude towards injection.

In situations with pressure competition, we show that there is no unambiguous baseline injection scenario for the case of no collaboration, only for the grand coalition implying full collaboration. Agents will affect each other by their injection choices, which adds an additional significant element of uncertainty to the variability due to unknown geological reservoir properties. We propose two stochastic physics-informed belief distribution models for the joint actions of all agents under competition. This allows a direct quantitative comparison in terms of stochastic preference relations of the alternatives cooperation and full or partial competition. The belief distributions exhibit more variability than the distribution for total CO$_2$ injection under collaboration. This is due to both intrinsic added uncertainty from unknown human behavior, and lack of information of the optimal injection scenarios under different man-made physical constraints due to pressure communication. While the amount of information is highly specific to any given storage project and one should avoid strong generalizations between projects, it is likely that the collaborative scenario exhibits significantly less variability, and remains the natural choice for risk-averse agents.

\section*{Declaration of Competing Interest}
The authors declare that they have no known competing financial interests or personal relationships that could have appeared to influence the work reported in this paper.

\section*{Acknowledgments}
This work was funded by the Norwegian Research Council grant 336294, Expansion of Resources
for CO2 Storage on the Horda Platform (ExpReCCS).

\appendix
\section{Equally or more attractive allocations}
\label{sec_appendix_redist}

Let the baseline quantile for agent $i$ be denoted $c_i$. Any allocation $(d_i, r_i)$ that satisfies 
\begin{equation}
\label{eq:allocation_ineq}
d_i + a_{\alpha}r_i  \geq c_i, \quad r_i \geq 0 \quad 
\sum_{j} d_j = E(Y), \quad 
\sum_{j} r_j = 1,
\end{equation}
is a feasible allocation that is more attractive to the agents than their baseline allocations.
For one such feasible allocation $(d_i', r_i')$ with quantiles $c_i'$,
\begin{equation}
d_i' + a_{\alpha}r_i'  = c_i', \quad r_i' \geq 0 \quad 
\sum_{j} d_j' = E(Y), \quad 
\sum_{j} r_j' = 1,
\end{equation}
if $r'_i > 0$ for at least one $i$, and assuming $\alpha_i$ being the same for all $i$, then the allocation $(d_i, r_i) = (d'_i + \Delta d_i, r'_i+\Delta r_i)$ satisfying $0 = \sum_j \Delta d_j = \sum_j \Delta r_j$ is also feasible with equal preferences, i.e., the agents are equally happy with $(d'_i, r'_i)$ and $(d'_i + \Delta d_i, r'_i+\Delta r_i)$. This implies that in general, an allocation leading to a certain degree of happiness is not unique.

For a feasible (attractive) allocation we can reformulate~\eqref{eq:allocation_ineq} as an equality:
\begin{equation}
\label{eq:allocation_eq}
d_i + a_{\alpha}r_i  = c_i + k_i, \quad r_i \geq 0 \quad 
\sum_{j} d_j = E(Y), \quad 
\sum_{j} r_j = 1,
\end{equation}
with $k_i\geq 0$. Summing over the agents in the first equality yields
\[
\sum_{i}d_i + a_{\alpha}r_i  = \sum_{i} c_i + k_i
\]
\[
E(Y) + a_{\alpha} = E(Y) + F_Y^{-1}(\alpha)-E(Y) = q^Y_{\alpha} = \sum_i c_i + k_i.
\]
Since this implies
\[
\sum_{i}k_i = q_{\alpha}^{Y} - \sum_{i} q_{\alpha}^{Y_i},
\]
there is only a total  positive surplus to distribute if 
\begin{equation}
\label{eq:surplus_cond_app}
q_{\alpha}^{Y} > \sum_{i} q_{\alpha}^{Y_i}.
\end{equation}

\section{Identical distribution of the individual injections}
\subsection{Independent distributions}
\label{sec:appendix_IID}
If $Y_i$ follows the same distribution for all $i=1,\dots, \Na$, then
\begin{equation}
\begin{aligned}
V(Y) &= \Na V(Y_i),\\
E(Y) &= \Na E(Y_i).
\end{aligned}
\end{equation}
Using the above expressions, the variance of the proportional allocation~\eqref{eq:risk-sharing-phys-prop} is given by:
\[
V\mleft(Y_i^{\text{prop}}\mright) = \mleft(\frac{E(Y_i)}{E(Y)}\mright)^2 V(Y) = \frac{1}{\Na^2} \Na V(Y_i)
 = \frac{1}{\Na} V(Y_i).
\]
Clearly, $V\mleft(Y_i^{\text{prop}}\mright) < V(Y_i)$ whenever $\Na \geq 2$. It then follows that:
\begin{equation}
\begin{aligned}
q_{\alpha_i}^{Y_i^{\text{prop}}} &> q_{\alpha_i}^{Y_i}, \quad \mbox{if } 0 \leq \alpha_i < 0.5,\\
q_{\alpha_i}^{Y_i^{\text{prop}}} &< q_{\alpha_i}^{Y_i}, \quad \mbox{if } 0.5 \leq \alpha_i \leq 1 .
\end{aligned}
\end{equation}
Hence, all risk averse agents will want to collaborate.

\subsection{Fully correlated injections}
\label{sec:appendix_ID_pos_corr}
In the case that the $Y_i$ follow identical distributions, but are perfectly correlated so that $Cov(Y_i, Y_j) = V(Y_i)$, for all $i,j$, we have
\[
V(Y_i^{\text{prop}}) = 
\frac{1}{\Na^2} \left( \Na V(Y_i) + \frac{\Na (\Na -1)}{2} V(Y_i)
\right)
 = \frac{\Na + 1}{2 \Na} V(Y_i)
\]
For $\Na\geq 2$,
\[
\frac{1}{\Na} V(Y_i)
<
\frac{\Na + 1}{2 \Na} V(Y_i) < V(Y_i),
\]
so risk averse agents will still want to collaborate with positively correlated distributions.


\subsection{Subset of small variability injection}
\label{sec:appendix_small_var}
Let all $Y_i$ be independent with identical expectations, and let $J$ be a subset of the indices $1,\dots, \Na$ for which $V(Y_j)=\epsilon V(Y_i)$ for all $j \in J$ and $i \notin J$. Then,
\[
V(Y) = |J| \epsilon V(Y_i) + (\Na - |J|)V(Y_i),
\]
for any $i \notin J$ and $|J|$ denoting the number of elements in $J$. Hence,
\[ 
V\mleft(Y_i^{\text{prop}}\mright) = 
\frac{1}{\Na^2} V(Y) = \frac{|J| \epsilon + (\Na - |J|)}{\Na^2}V(Y_i) .
\]
Thus, a risk averse $a_j$ will want to collaborate if
\[
\epsilon V(Y_i) = V(Y_j) > V\mleft(Y_i^{\text{prop}}\mright)
= \frac{|J| \epsilon + (\Na - |J|)}{\Na^2}V(Y_i). 
\]
Canceling $V(Y_i)\neq 0$, this holds when
\[
\epsilon > \frac{\Na - |J|}{\Na^2 - |J|}.
\]
When $|J|=1$,
\[
\epsilon > \frac{\Na - 1}{\Na^2 - 1} = \frac{1}{\Na + 1}.
\]



\printbibliography


\end{document}